\documentstyle[12pt]{article}
\renewcommand{\slash}[1]{/\kern-7pt #1}
\renewcommand{\theequation}{\thesection.\arabic{equation}}

\def\N1{N=1 supersymmetric }
\def\YM{Yang-Mills }
\def\HV{`t~Hooft-Veltman }
\def\FDH{four-dimensional helicity }
\def\DR{dimensional reduction }
\def\CDR{conventional }
\def\1PI{one-particle irreducible }
\def\SWI{supersymmetry Ward identities}
\def\NLO{next-to-leading order}
\def\AP{Altarelli-Parisi }
\def\IR{infrared }
\def\UV{ultraviolet }
\def\tr{{\rm Tr}}
\def\d{{\rm d}}

\def\i{{\rm i}}
\def\A{{\cal A}}
\def\F{{\cal F}}
\def\I{{\cal I}}
\def\M{{\cal M}}
\def\S{{\cal S}}
\def\ds{\displaystyle}
\def\ra{\rightarrow}
\def\ms{$\overline{{\rm MS}}$}
\def\lQCD{$\Lambda_{{\rm QCD}}$}
\def\l{\langle}
\def\r{\rangle}
\def\vspaceinarray{\nonumber ~&~&~\\}

\begin{document}

\begin{titlepage}
\vspace*{-2cm}
\begin{flushright}
ETH-TH/93-11\\
May 5 1992 \\
\end{flushright}
\vskip .5in
\begin{center}
{\Large\bf
One-loop helicity amplitudes for all ${\bf 2\ra2}$ processes
in QCD and \N1  \YM  theory}
\footnote{Work supported in part by the Schweizerischer
Nationalfonds}\\
\vskip 1cm
{\large Zoltan Kunszt, Adrian Signer and Zolt\'an  Tr\'ocs\'anyi} \\
\vskip 0.2cm
Theoretical Physics, ETH, \\
Z\"urich, Switzerland  \\
\vskip 1cm
\end{center}

\begin{abstract}
\noindent One-loop corrections to the helicity amplitudes of
all $2\ra 2$ subprocesses are calculated in QCD and in \N1 \YM
theory using two versions of dimensional regularization: the
\HV scheme and dimensional reduction.  Studying the structure
of the soft and collinear singularities, we found universal
transition rules for the squared matrix element which can be
used to translate the results obtained in these schemes to the
results valid in the conventional dimensional regularization
scheme. With explicit calculation it is demonstrated that the
one-loop helicity amplitudes of the $2\ra 2$ subprocesses
calculated using dimensional reduction in the \N1 $SU(N)$ gauge
theory  respect the \SWI.  Our transition rules can also be
used to calculate the \NLO\ \AP kernels in the dimensional
reduction scheme when they satisfy \SWI\ as well.
\end{abstract}
\end{titlepage}
\setcounter{footnote}{0}

\bigskip

\section{Introduction}

This paper discusses the consistency of various versions of
the dimensional regularization schemes for regulating both
ultraviolet and infrared divergences in calculations of
infrared safe quantities for hadron-hadron collisions
at \NLO\ in QCD and in \N1 \YM theory. The discussion is based
upon the calculation of the one-loop radiative corrections to the
helicity amplitudes of all $2\ra 2$ parton scattering processes.
 The motivation for this study is a recent
paper by Bern and Kosower \cite{Ber92} in which the authors
worked out string-theory-based technique for evaluating
multi-gluon amplitudes in one-loop order. The power of the
method is demonstrated clearly by the recent calculation of the
one-loop corrections to five gluon helicity amplitudes using
the string-theory-based technology \cite{Ber93}.
The field theory interpretation of the new results
suggested that the use of background Feynman gauge, the helicity
method and  a new version of dimensional regularization, the
so called \FDH scheme, results in great technical advantages
also in the standard field theory calculations.

In perturbative QCD, with the evaluation of the loop corrections
one does not obtain yet physical cross-sections. The Bremsstrahlung
contributions have to be added as well. In particular, finite
hard scattering cross-sections are obtained only (after trivial
ultraviolet renormalization) by cancelling soft and collinear
singular terms between loop corrections and Bremsstrahlung
contributions, and by subtracting the initial-state collinear
singularities. In ref.\ \cite{Kun92}, the singular terms appearing
in this procedure have been worked out analytically using
the \CDR dimensional regularization scheme. The analytic expression
obtained in ref.\ \cite{Kun92} for the soft and collinear contributions
have universal, process independent character, but they depend
on the regularization scheme used. The sensitivity to the
regularization scheme, however, are also exhibited analytically,
therefore, the result of ref.\ \cite{Kun92} can easily be transformed
to the various versions of the dimensional regularization schemes.

Physical cross sections of infrared safe quantities are obtained
by folding the finite hard-scattering cross section with parton
densities. In a \NLO\ calculation the $Q^2$-evolution of the
parton densities has to be carried out with the \NLO\ \AP kernels
which also depend on the regularization, factorization and
ultraviolet-subtraction schemes. In the conventional \ms\ factorization
scheme the \AP  kernels have been calculated by Curci, Furmanski
and Petronzio \cite{Cur80}. Changing the regularization scheme will
generate  changes in the \NLO\ \AP kernels, in the hard-scattering
cross sections and in the value of \lQCD, but the physical
cross section has to remain the same. Therefore, calculating the
hard-scattering cross sections in different schemes, we can identify
the  universal transition terms which define the transformation of the
\NLO\ \AP kernel from one scheme to another.

Recently, it has become clear that with appropriately modified
dimensional regularization schemes the advantages of the helicity
method can be maintained also for loop calculations \cite{Ber92,Gie92}.
Three modifications have been proposed: the \HV scheme, regularization
by dimensional reduction and the so called \FDH scheme\footnote{The
\FDH scheme was proposed in the context of the string-theory derivation
of the one-loop corrections to the four-gluon helicity amplitudes
\cite{Ber92}. According to our knowledge, no operational definition
of this regularization has been defined in the standard field theory
framework.}. Changing the regularization scheme, the soft and
collinear singular contributions will change but this change
has universal character. Therefore, it will be sufficient to find
the changes generated in the hard scattering cross sections
only for few processes. Then one knows the transition
functions for all other processes. Deriving the transition terms
has great practical importance because the phenomenological fit
of the parton densities and their $Q^2$-evolution is worked out
in the \CDR regularization scheme. Therefore, if we calculate
the \NLO\ corrections in some other schemes, we must be able
to transform the result into the conventional scheme.

In this paper, with the explicit calculation of the \NLO\
corrections to all $2\ra 2$ parton-parton scattering amplitudes, we
determine the transition rules necessary to translate the results
obtained in the \HV and \DR schemes into those in the \CDR scheme.
The rules are process independent, therefore, they can be applied
to other processes. For example, calculating the \NLO\ corrections
to the $2\ra 3$ scattering amplitudes, one can use such a regularization
scheme that is consistent with the application of the helicity method
and supersymmetry (dimensional reduction) and then one can easily
translate the result into the phenomenologically relevant \CDR scheme.

It has been noted long time ago that in calculating
scattering amplitudes of massless partons in perturbative
QCD, supersymmetry can be used as a technical tool to
check and simplify the calculation. The reason for this is
that if we change the color representation of quarks to the
adjoint representation of a Majorana fermion we obtain the \N1 \YM
theory in the Wess-Zumino gauge. If we use \DR which respects
supersymmetry, then we can test \SWI\ between the amplitudes of
various subprocesses with different number of external fermions and
bosons.  These relations, which are valid in all order in perturbation
theory, provide us a very significant check on the correctness
of the calculation. One can also turn the argument around and use
the \SWI\ to obtain gluonic amplitudes from the fermionic ones
correcting for the differences in the contribution of fermion loops.

We carried out all the calculations presented in this
paper using standard Feynman diagram method, but using
all the advantages of the helicity method, the background Feynman
gauge and the \HV and \DR regularization schemes. The string theory
derivation of the gluonic amplitudes in ref.\ \cite{Ber92}
suggests that significant technical advantages
can be achieved in addition with cleverly combining class of
Feynman diagrams and reducing tensor integrals to scalar integrals.
We did not make attempts to simplify our calculation in these
directions. However, we used color subamplitudes \cite{Ber87}
and the method of
ref.\ \cite{Lam69} to reduce the tensor loop integrals to scalar
integrals which is easily adaptable to calculations in massless QCD.
This part of our field theory calculation still cannot match the
simplicity achieved by the string-based technique.

The organization of our paper is as follows. In sect.\ 2, we describe
the various versions of dimensional regularization briefly. In sects.\
3 and 4, we give an overview of the techniques which enter our
calculations: the helicity method and color decomposition of the
amplitudes, the reduction of tensor integrals and use of background
field method. These technicalities have been discussed in much
details elsewhere in the literature, therefore, our presentation
is constrained to setting the notation. Sect.\ 5 contains the one-loop
color subamplitudes in QCD which constitutes one of our main results.
We present the results in the unphysical region where all kinematic
variables are negative which may become useful in the future.
We also calculate the shift in the strong coupling constant caused
by using different regularization schemes. In sect.\ 6, we
calculate the \NLO\ loop-correction to the square of the matrix
element in order that we can compare our results to that of Ellis and
Sexton \cite{Ell86}. We establish transition rules among the
one-loop amplitudes obtained using different versions of dimensional
regularization. Sect.\ 7 presents the one-loop color subamplitudes
in the supersymmetric limit of QCD, \N1 \YM theory, and elaborates on
the usefulness of supersymmetry in QCD calculations.
In order to find the complete
transition rules among the various regularization schemes, in sect.\
8, we discuss the difference among the Bremsstrahlung contributions
obtained using different regulators. In sect.\ 9, we shall spell out
how the hard-scattering cross section changes with changing the
regularization scheme. In sect.\ 10, we shall discuss how the change
in the hard-scattering cross section due to the use of different
regularization schemes can be compensated by a change in the parton
density functions. In this way, we can present some consistency checks
on the correctness and process independent property of our transition
rules. Finally, sect.\ 11 will summarize the main results.

\section{Regularization schemes within dimensional regularization}
\setcounter{equation}{0}

In massless QCD, the \CDR dimensional regularization is an attractive
scheme because it simultaneously regulates \IR and \UV divergences, is
manifestly gauge invariant, consistent with unitarity and simple to
implement.

When defining the different versions of dimensional regularization
both for loop and phase space integrals, it is useful to distinguish
two class of particles: observed and unobserved ones. Unobserved
particles are those virtual ones which circulate in internal loops as
well as those which are external but soft or collinear with other
external particles. All the rest are observed particles.
Unitarity demands that unobserved particles are treated uniformly.

The most important ingredient of dimensional regularization is the
continuation of the momenta of the unobserved particles into
$d\ne 4$ dimensions, thus rendering the integrals over these momenta
finite. Having this done, one is left with a lot of freedom how to
treat the momenta of observed particles and the polarization vectors
of all particles. The different choices lead to different versions of
dimensional regularization.

The original choice made by \HV \cite{'tH72} was to continue both the
momenta and the helicities of unobserved particles into $d$ dimensions,
while leaving the momenta and helicities of observed particles in four
dimensions.\footnote{As a matter of fact, the original proposal was
meant for \UV regularization of loop integrals only. The possibility
of using it also for regulating \IR singularities was initiated in
ref.\ \cite{Gas73}. For a more detailed recent application see
ref.\ \cite{Gie92}.}

There is a third version of dimensional regularization, the \DR
\cite{Sie79}, widely used in connection with supersymmetric theories.
This technique consist of continuing the virtual momenta of loop
integrals into dimensions $d<4$, but keeping all polarization
vectors in four dimensions. Because of $d<4$, the method is not
directly applicable for regulating \IR divergent phase space integrals.
It is essential that $d<4$ because the four-dimensional vector field
is split into a $d$-dimensional vector plus a field which transforms
as ($4-d$)-dimensional scalars under gauge transformation \cite{Cap80}.
It can be shown, however, that the use of this Lagrangian is
equivalent to using the original Lagrangian without the splitting of
the gauge field and perform the algebra in four dimensions. The only
subtle point which is a reminiscent of the splitting is that one has
to distinguish between four-dimensional metric tensors coming from the
Feynman rules and $d$-dimensional ones emerging from momentum integrals
with more than one loop momentum in the numerator. In this way, gauge
invariance is maintained which has been checked explicitly up to two
loops \cite{Cap80}. We remark that the operational definition of \DR
is not complete in complicated cases (see e.g.,  refs.\
\cite{Sie80,Avd83}). However, this incompleteness does not affect our
calculation.

In supersymmetric theories, there is a different formulation of \DR
when supergraph technique is applied. In this approach all algebra is
performed in four dimensions until only final scalar integrals are
left in which the continuation of the loop momenta is performed. In
ref.\ \cite{Cap80}, for relatively simple calculations the equivalence
of the two approaches has been pointed out.

Inspired by string-based rules for calculating loop amplitudes in
gauge theories, in ref.\ \cite{Ber92} a new version of dimensional
regularization, the \FDH scheme has been proposed. The operational
definition in the string calculation is such that only the momenta of
unobserved particles are continued into $d$ dimensions, all
polarization vectors and the momenta of observed particles are kept in
four dimensions. The meaning of such a definition in field theory is
obvious in most of the
algebra except for the reduction of tensor loop integrals. In the
string-based calculation, the loop momentum is integrated out at the
string level which implies that the reduction of tensor integrals has
been implicitly achieved  in four dimensions. Therefore, in the field
theory version of the \FDH scheme, one may try to set $d=4$ in the
algebraic part of tensor loop integrals and treat the remaining
scalar loop integral in $d$ dimensions. In the present paper, we do
not pursue this possibility any further.

The requirements that a certain regularization scheme is a consistent
regulator in a gauge field theory are that it has to respect gauge
invariance and unitarity. Gauge invariance is known to be maintained by
the conventional, \HV and \DR schemes. Unitarity is obviously
maintained if the momenta and helicities of all particles are
continued into $d$ dimensions, therefore, the \CDR scheme
is considered a consistent scheme in massless QCD. The other three
versions of dimensional regularization do not treat momenta and
helicities equally, therefore, it is not obvious that unitarity is
respected by these schemes.

In axial gauge, the collinear singularities come from self-energy
contributions on external lines only \cite{Cur80}, therefore, it is
expected that unitarity is preserved in the \HV scheme. In the case of
\DR and \FDH scheme, we do not know such a general argument.

\section{The helicity method and color partial amplitudes}
\setcounter{equation}{0}

We use the helicity method in its simplest \cite{Xu87}, crossing
symmetric version \cite{Gun85}. In order to fix our notation,
we present the definitions for the spinor algebra here. For more
details, the reader is referred to Appendix A of ref.\ \cite{Man91}.

Let $\psi(p)$ be a four-dimensional spinor satisfying the massless
Dirac equation:
\begin{equation}
\slash{p}\psi(p)=0,\;p^2=0.
\end{equation}
We define the two helicity states of $\psi(p)$ by the two chiral
projections
\begin{equation}
\psi^\pm(p)=\frac{1}{2}(1\pm\gamma_5)\psi(p)=(\psi^\mp(p))^c.
\end{equation}
The second equation is a conventional choice of relative phase between
opposite helicity spinors fixed by the properties under charge
conjugation $C$,
\begin{equation}
(\psi(p))^c=C(\psi(p))^*,\; C\gamma_\mu^*C^{-1}=\gamma_\mu.
\end{equation}
Following ref.\ \cite{Xu87,Gun85}, we introduce a new notation
\begin{equation}
|p\pm\r = \psi_\pm(p),\; \l p\pm| = \overline{\psi_\pm(p)},
\end{equation}
\begin{equation}
\l pq \r = \l p-|p+ \r,\; [pq] = \l p+|p- \r.
\end{equation}
With this notation the normalization of the spinor is expressed as
\begin{equation}
\l p\pm|\gamma_\mu|p\pm \r = 2p_\mu,
\end{equation}
and we have the useful property
\begin{equation}
|p \pm \r\l p\pm| = \frac{1}{2}(1\pm\gamma_5)\slash{p}.
\end{equation}

The polarization vector of an outgoing massless vector of momentum
$p$ is defined as
\begin{equation}
\varepsilon^\pm(p,k)=
\pm\frac{\l p\pm|\gamma_\mu|k\pm\r}{\sqrt{2}\l k\mp|p\pm\r}
\end{equation}
where $k$ is an arbitrary (reference) momentum.
Then we find that the usual requirements for the polarization vectors,
\begin{equation}
\varepsilon^+(p,k)\cdot\varepsilon^+(p,k) = 0,
\end{equation}
\begin{equation}
\varepsilon^+(p,k)\cdot\varepsilon^-(p,k) = -1
\end{equation}
are satisfied.

As mentioned in the introduction, we also make use of the color
subamplitudes \cite{Ber87} which give gauge invariant decomposition
in color space. For the tree and one-loop four-point amplitudes we
find\footnote{We use the same notation as in ref.\
\cite{Ell86} to distinguish the different processes and take all
particles outgoing. We note that in the case of gluon-gluon
scattering, our decomposition differs formally from that given in
ref.\ \cite{Ber92}. The difference is the omission of the term
proportional to $\tr\,T^a=0$.}:

\newpage
\begin{eqnarray}
\label{treeA}
\A_4^{\rm tree}(\bar{q},\bar{Q};Q,q)&=&
g^2\left(\delta_{i_1i_3}\delta_{i_2i_4}
-\frac{1}{N_c}\delta_{i_1i_4}\delta_{i_2i_3}\right)
a_{4;0}(1,2;3,4).\\
\vspaceinarray
\label{treeB}
\A_4^{\rm tree}(\bar{q},\bar{q};q,q)&=&
g^2\sum_{\sigma\in S_2}
\left(\delta_{i_1\sigma(3)}\delta_{i_2\sigma(4)}
-\frac{1}{N_c}
\delta_{i_1\sigma(4)}\delta_{i_2\sigma(3)}\right)\\ \nonumber
&&~~~~~~~~~~\times
b_{4;0}(1,2;\sigma(3),\sigma(4)).\\
\vspaceinarray
\label{treeC}
\A_4^{\rm tree}(g,g;q,\bar{q})&=&
g^2\sum_{\sigma\in S_2}
\left(T^{a_{\sigma(1)}}T^{a_{\sigma(2)}}\right)_{i_3i_4}
c_{4;0}(\sigma(1),\sigma(2);3,4).\\
\vspaceinarray
\label{treeD}
A_4^{\rm tree}(g,g,g,g)&=&
g^2\sum_{\sigma\in S_4/Z_4}
\tr(T^{a_{\sigma(1)}}T^{a_{\sigma(2)}}
T^{a_{\sigma(3)}}T^{a_{\sigma(4)}})\\ \nonumber
&&~~~~~~~~~~\times
d_{4;0}\left(\sigma(1),\sigma(2),\sigma(3),\sigma(4)\right).\\
\vspaceinarray
\label{oneloopA}
\A_4^{\rm 1-loop}(\bar{q},\bar{Q};Q,q)&=&
g^4\left[\left(\delta_{i_1i_3}\delta_{i_2i_4}
-\frac{1}{N_c}\delta_{i_1i_4}\delta_{i_2i_3}\right)
a_{4;1}(1,2;3,4)\right. \\ \nonumber
&&~~~~~~\left.
+\delta_{i_1i_3}\delta_{i_2i_4}a_{4;2}(1,2;3,4) \right].\\
\vspaceinarray
\label{oneloopB}
\A_4^{\rm 1-loop}(\bar{q},\bar{q};q,q)&=&
g^4\sum_{\sigma\in S_2}\left[
\left(\delta_{i_1\sigma(3)}\delta_{i_2\sigma(4)}
-\frac{1}{N_c}
\delta_{i_1\sigma(4)}\delta_{i_2\sigma(3)}\right)\right.\\ \nonumber
&&~~~~~~~~~~~~\times
b_{4;1}(1,2;\sigma(1),\sigma(2))\\ \nonumber
&&~~~~~~~~~\left.
+\delta_{i_1\sigma(3)}\delta_{i_2\sigma(4)}
b_{4;2}(1,2;\sigma(3),\sigma(4)) \right].\\
\vspaceinarray
\label{oneloopC}
\A_4^{\rm 1-loop}(g,g;q,\bar{q})&=&
\left.g^4\right[\sum_{\sigma\in S_2}
\left(T^{a_{\sigma(1)}}T^{a_{\sigma(2)}}\right)_{i_3i_4}
c_{4;1}(\sigma(1),\sigma(2);3,4) \\ \nonumber
&&~~~~~~~\left.
+\delta_{i_3i_4} \delta_{a_1a_2}c_{4;2}(1,2;3,4)\right].\\
\vspaceinarray
\label{oneloopD}
\A_4^{\rm 1-loop}(g,g,g,g)&=&
\left.g^4\right[\sum_{\sigma\in S_4/Z_4}
\tr(T^{a_{\sigma(1)}}T^{a_{\sigma(2)}}
T^{a_{\sigma(3)}}T^{a_{\sigma(4)}})\\ \nonumber
&&~~~~~~~~~~\times
d_{4;1}\left(\sigma(1),\sigma(2),\sigma(3),\sigma(4)\right) \\
\nonumber &&~~~~+\sum_{\sigma\in S_4/Z_2^3}
\tr(T^{a_{\sigma(1)}}T^{a_{\sigma(2)}})
\tr(T^{a_{\sigma(3)}}T^{a_{\sigma(4)}})\\ \nonumber
&&~~~~~~~~~~\times
\left.d_{4;2}\left(\sigma(1),\sigma(2);\sigma(3),\sigma(4)\right)\right].
\end{eqnarray}
In these equations, $S_n$ is the permutation group of $n$ elements,
$Z_n$ is the cyclic group of $n$ elements, i.e., $S_n/Z_n$ means the
group of permutations of $n$ elements with cyclic permutations removed.

The helicity subamplitudes for $n$-gluon processes enjoy a
number of useful properties \cite{Ber92} which simplify the actual
calculation (or alternatively, can be used as checks of the results).
The amplitudes of the other processes posses much less symmetry.
However, useful relations can be discovered in the supersymmetric
limit (see sect.\ 7).

The Feynman diagrams for quark-quark, quark-gluon and gluon-gluon
scattering have been presented in ref.\ \cite{Ell86}; we do not repeat
them in our paper. The diagrams that enter the calculation of a given
partial amplitude can be found by imposing the condition that they
have the proper color structure.  In the sect.\ 4, we describe how
to compute a one-loop Feynman diagram.
The external legs of the diagrams to be computed are on shell, i.e., for
massless partons the momentum squared of the external legs is zero.
In dimensional regularization a loop insertion on a particle line is
proportional to the momentum squared of the line. This means that those
diagrams which have a loop insertion on an external line are zero in
dimensional regularization. Motivated by the results of ref.\
\cite{Dun92}, we shall use background Feynman gauge. The
necessary Feynman rules can be found in the standard reference
\cite{Abb81}.

\section{Calculation of one-loop Feynman diagrams}
\setcounter{equation}{0}

One of the methods we used for calculating \1PI Feynman diagrams is
as described in ref.\ \cite{Lam69}. In this section, we recall the
basic ingredients of the method and give all necessary formul\ae\
for calculating four-parton processes explicitly.

The general algebraic expression for a one-loop \1PI Feynman diagram
with $N$ propagators in $d=2\omega$ ($\omega=2-\varepsilon$)
dimensions is
\begin{equation}
\M=\mu^{2\varepsilon}\int\frac{\d^{2\omega}\ell}{(2\pi)^{2\omega}}\,
\frac{N(q,p)}{\prod_{r=1}^N(q_r^2-m_r^2+\i\eta)},
\end{equation}
where $\ell$ is the loop momentum, $p_i$ are the external outgoing
momenta, $q_r$ are the momenta of the propagators and $N(q,p)$ is the
numerator which receives its contributions from vertices and numerators
of propagators and contains the dimensionless coupling.\footnote{A
factor of $\mu^{(N-2)\varepsilon}$ is omitted. When a cross section is
calculated this factor appears in the corresponding tree diagrams, as
well. Therefore, for the cross section it means an irrelevant overall
factor which disappears at the end of the calculation when physical
dimension is considered.} After Feynman parametrization the loop integral
can be performed to give
\begin{equation}
\label{parametrized}
\M=(-1)^N\frac{\i}{(4\pi)^2}(4\pi\mu^2)^\varepsilon
\sum_{k=0}\Gamma(N-\omega-k)\int\prod_{r=1}^N\d x_r\,
\frac{N_k(J,p)}{D(x,p)^{N-\omega-k}}
\delta\left(\sum_{r=1}^N x_r-1\right).
\end{equation}

Cutting two of the propagators in the loop produces two tree diagrams
--- a two-tree.  A double cut can be performed $\ds N\choose\ds 2$
different ways.  The set of two-trees will be denoted by $T_2$.
With this notation and for massless particles in the propagators
\begin{equation}
D(x,p)=-\sum_{T_2}(x_{c_i}x_{c_j})\left(\sum_c p_c\right)^2,
\end{equation}
where $x_{c_i}$, $x_{c_j}$ are the Feynman parameters of the two cut
lines and the sum over $c$ is the sum of the external momenta belonging
to one of the trees. In the numerator in eq.\ (\ref{parametrized}),
$N_0(J,p)=N(J,p)$ and
\begin{equation}
J_r=\sum_{T_2(r)}x_{c_i}\left(\sum_c p_c\right).
\end{equation}
In this equation one of the cut lines is $r$, while $x_{c_i}$ is the
Feynman parameter of the other cut line. In the sum over $c$, the sum of
the external momenta belonging to the tree to which $q_r$ flows is to
be taken. $N_k(J,p)$ for $k>0$ is obtained from $N_0(J,p)$ by contracting
$k$ pairs of $J$'s in all possible ways\footnote{This contraction is
done in $2\omega$
dimensions both in \HV and \DR schemes.} and summing over all such
contractions. If no contraction is possible then $N_k=0$. Contraction of
$J$'s means
\begin{equation}
J_r^\mu J_s^\nu\ra -\frac{1}{2}g_{(2\omega)}^{\mu\nu}.
\end{equation}
This rule is valid in this simple form only if the propagators in the
loop constitute a continuos flow of the momentum which for a one-loop
diagram can always be arranged.

We give formula (\ref{parametrized}) for the cases that enter our
calculation ($N=2,3,4$) together with the necessary integrals
explicitly in Appendix A.

As a second method, we calculated the one-loop tensor integrals using
the standard Passarino-Veltman type scheme \cite{Pas79}.

We note that the tree level diagrams with one-loop renormalization
insertions have also to be taken into account. The \ms\ \UV counterterm
for the $n$-point scattering amplitude at the one-loop level is well
known:
\begin{equation}
(n-2)\left[-\frac{1}{2}\beta_0\left(\frac{g}{4\pi}\right)^2
\frac{(4\pi)^\varepsilon}{\varepsilon\Gamma(1-\varepsilon)}
\A_n^{{\rm tree}}(1,\ldots,n)\right],
\end{equation}
where $\beta_0=(11N_c-2N_f)/3$ and
$\A_n^{{\rm tree}}$ is the four-dimensional tree-level amplitude.

\section{One-loop color subamplitudes in QCD}
\setcounter{equation}{0}

In this section, we present the results obtained for the QCD
color subamplitudes. First, we list the amplitudes in the \HV
scheme. Next, we give the differences of the amplitudes obtained in
the \HV and \DR schemes. Finally, we calculate
$\Lambda_{\overline{\rm MS}}^{{\rm DR}}$ for \DR in terms of
$\Lambda_{\overline{\rm MS}}$ of \CDR dimensional
regularization.\footnote{$\Lambda_{\overline{\rm MS}}$
in the \HV and \CDR schemes are identical.}

In order to check the following results, we calculated the nontrivial
integrals in two different ways (one is described in Appendix A)
and checked gauge invariance in all external gluon legs using
longitudinal gluons. The amplitudes for gluon-gluon scattering without
the fermionic contribution have already been calculated in ref.\
\cite{Ber92}. Our results are in complete agreement with the published
ones.

\subsection{Results in the \HV scheme}

In the following, we constrain the presentation of the results to the
minimal information from which after trivial algebra --- i.e., using
parity transformation or the cyclic property of four-gluon amplitudes
--- any helicity amplitude can be obtained.
To calculate the square of the matrix element for process B, the same
partial amplitudes are to be used as for process A, only the color and
helicity sums differ. Therefore, the subamplitudes for processes A and
B are not written separately. The $d_{4;2}$ subamplitudes will not
play any role in the rest of this paper. They can be obtained from the
from the $N_f$ independent part of the $d_{4;1}$ amplitudes using the
decoupling equations of ref.\ \cite{Ber92}, where they were published,
therefore, we do not repeat them here. We remind the reader that
according to our convention all particles are outgoing.\footnote{We
have checked the results for Processes A and C in the \HV scheme
using Passarino-Veltman type reduction of tensor integrals.}

We start our description of the results with those amplitudes which are
finite in four dimensions (the corresponding tree amplitudes are vanishing).

\bigskip
Process A:

\bigskip
The amplitudes which vanish at tree level vanish at one-loop level as
well.

\bigskip
Process C:

\bigskip
\begin{eqnarray}
c^{{\rm HV}}_{4;1}(+,+;+,-)&=&
-\frac{\i}{48\pi^2}\frac{\l 34\r[12][13][23]}{s_{12}s_{14}}
\left[\frac{3}{2}\frac{N_c^2+1}{N_c}
+(N_c-N_f)\frac{s_{14}}{s_{12}}\right].\\
\vspaceinarray
c^{{\rm HV}}_{4;1}(-,-;+,-)&=&
\frac{\i}{48\pi^2}\frac{\l 12\r\l 14\r\l 24\r [34]}{s_{12}s_{14}}
\left[\frac{3}{2}\frac{N_c^2+1}{N_c}
+(N_c-N_f)\frac{s_{14}}{s_{12}}\right].\\
\vspaceinarray
c^{{\rm HV}}_{4;2}(+,+;+,-)&=&c^{{\rm HV}}_{4;2}(-,-;+,-)=0.
\end{eqnarray}

\bigskip
Process D:

\bigskip
\begin{eqnarray}
d^{{\rm HV}}_{4;1}(+,+,+,+)&=&\frac{\i}{48\pi^2}\left(N_c-N_f\right)
\frac{s_{12}s_{14}}{\l 12\r\l 23\r\l 34\r\l 41\r}.\\
\vspaceinarray
d^{{\rm HV}}_{4;1}(-,+,+,+)&=&\frac{\i}{48\pi^2}\left(N_c-N_f\right)
\frac{[24]^2}{[12]\l 23\r\l 34\r[41]}(s_{12}+s_{14}).\\
\end{eqnarray}

All the other partial amplitudes are divergent in four dimensions.
We present the
results in the unphysical region, where all kinematic variables are
negative, therefore, the arguments of the logarithms are positive.
The analytic continuation to the physical region can easily
be performed keeping in mind that the branch cut of the logarithm is
obtained by inserting the i$\eta$ associated with each
kinematic variable:
\begin{eqnarray}
\label{continue}
(-s_{ij})^{-\varepsilon}&\ra&
|s_{ij}|e^{-\i\pi\varepsilon\Theta(s_{ij})}, \\
\log(-s_{ij})&\ra&\log{|s_{ij}|}+\i\pi\Theta(s_{ij}),
\end{eqnarray}
where $\Theta(x)$ is the usual step function. The definition of
$\beta_0$, appearing in the following results, is given after eq.\
(4.6). One finds:

\newpage
Process A:

\bigskip

\begin{equation}
a^{{\rm HV}}_{4;1}(-,-;+,+)=c_\Gamma
a_{4;0}(-,-;+,+)F^{--}_{a;1}(\varepsilon, s_{12}, s_{13}, s_{14}),
\end{equation}
\begin{eqnarray}
\lefteqn{F^{--}_{a;1}(\varepsilon,s_{12},s_{13},s_{14})=}\\ \nonumber
&&\left(-\frac{\mu^2}{s_{14}}\right)^\varepsilon
\left\{N_c\left[-\frac{2}{\varepsilon^2}-\frac{3}{\varepsilon}
+\frac{11}{3\varepsilon}
-\frac{2}{\varepsilon}\log\frac{s_{14}}{s_{12}}
+\frac{13}{9}+\pi^2\right]
+N_f\left[-\frac{2}{3\varepsilon}
-\frac{10}{9}\right]\right. \\ \nonumber
&&~~~~~~~~~~~~
-\frac{1}{N_c}\left[-\frac{2}{\varepsilon^2}-\frac{3}{\varepsilon}
-\frac{2}{\varepsilon}\log\frac{s_{12}}{s_{13}}-8
+\frac{1}{2}\frac{s_{14}}{s_{12}}\left(1-\frac{s_{13}}{s_{12}}\right)
\left(\log^2\frac{s_{14}}{s_{12}}+\pi^2\right)\right. \\ \nonumber
&&~~~~~~~~~~~~~~~~~~\left.\left.
+\frac{s_{14}}{s_{12}}\log\frac{s_{14}}{s_{13}}\right]\right\}
-\frac{1}{\varepsilon}\beta_0.
\end{eqnarray}
\vspace*{10pt}
\begin{equation}
a^{{\rm HV}}_{4;1}(-,+;-,+)=c_\Gamma
a_{4;0}(-,+;-,+)F^{-+}_{a;1}(\varepsilon, s_{12}, s_{13}, s_{14}),
\end{equation}
\begin{eqnarray}
\lefteqn{F^{-+}_{a;1}(\varepsilon,s_{12},s_{13},s_{14})=}\\ \nonumber
&&\left(-\frac{\mu^2}{s_{14}}\right)^\varepsilon
\left\{N_c\left[-\frac{2}{\varepsilon^2}-\frac{3}{\varepsilon}
+\frac{11}{3\varepsilon}
-\frac{2}{\varepsilon}\log\frac{s_{14}}{s_{12}}
+\frac{13}{9}+\pi^2\right]\right.
+N_f\left[-\frac{2}{3\varepsilon}-\frac{10}{9}\right] \\ \nonumber
&&~~~~~~~~~~~~
-\frac{1}{N_c}\left[-\frac{2}{\varepsilon^2}-\frac{3}{\varepsilon}
-\frac{2}{\varepsilon}\log\frac{s_{12}}{s_{13}}-8\right] \\ \nonumber
&&~~~~~~~~~~~~\left.
\left(N_c+\frac{1}{N_c}\right)\left[
\frac{1}{2}\frac{s_{14}}{s_{13}}\left(1-\frac{s_{12}}{s_{13}}\right)
\left(\log^2\frac{s_{14}}{s_{13}}+\pi^2\right)
+\frac{s_{14}}{s_{13}}\log\frac{s_{14}}{s_{12}}\right]\right\}
-\frac{1}{\varepsilon}\beta_0.
\end{eqnarray}
\vspace*{10pt}
\begin{equation}
a^{{\rm HV}}_{4;2}(-,-;+,+)=c_\Gamma
a_{4;0}(-,-;+,+)F^{--}_{a;2}(\varepsilon, s_{12}, s_{13}, s_{14}),
\end{equation}
\begin{eqnarray}
\lefteqn{F^{--}_{a;2}(\varepsilon,s_{12},s_{13},s_{14})=}\\ \nonumber
&&\left(-\frac{\mu^2}{s_{14}}\right)^\varepsilon
\frac{V}{N_c}\left[-\frac{2}{\varepsilon}\log\frac{s_{12}}{s_{13}}
+\frac{1}{2}\frac{s_{14}}{s_{12}}\left(1-\frac{s_{13}}{s_{12}}\right)
\left(\log^2\frac{s_{14}}{s_{13}}+\pi^2\right)
+\frac{s_{14}}{s_{12}}\log\frac{s_{14}}{s_{13}}\right].
\end{eqnarray}
\begin{equation}
a^{{\rm HV}}_{4;2}(-,+;-,+)=c_\Gamma
a_{4;0}(-,+;-,+)F^{-+}_{a;2}(\varepsilon, s_{12}, s_{13}, s_{14}),
\end{equation}
\begin{eqnarray}
\lefteqn{F^{-+}_{a;2}(\varepsilon,s_{12},s_{13},s_{14})=}\\ \nonumber
&&\left(-\frac{\mu^2}{s_{14}}\right)^\varepsilon
\frac{V}{N_c}\left[-\frac{2}{\varepsilon}\log\frac{s_{12}}{s_{13}}
-\frac{1}{2}\frac{s_{14}}{s_{13}}\left(1-\frac{s_{12}}{s_{13}}\right)
\left(\log^2\frac{s_{14}}{s_{12}}+\pi^2\right)
-\frac{s_{14}}{s_{13}}\log\frac{s_{14}}{s_{12}}\right].
\end{eqnarray}

\newpage
Process C:

\bigskip
\begin{equation}
c^{{\rm HV}}_{4;1}(-,+;+,-)=c_\Gamma
c_{4;0}(-,+;+,-)F^{-+}_{c;1}(\varepsilon, s_{12}, s_{13}, s_{14}),
\end{equation}
\begin{eqnarray}
\lefteqn{F^{-+}_{c;1}(\varepsilon,s_{12},s_{13},s_{14})=}\\ \nonumber
&&\left(-\frac{\mu^2}{s_{12}}\right)^\varepsilon
\left\{N_c\left[-\frac{3}{\varepsilon^2}-\frac{3}{2\varepsilon}
+\frac{2}{\varepsilon}\log\frac{s_{13}}{s_{12}}
-\frac{7}{2}+\frac{1}{2}\pi^2
-\frac{1}{2}\log^2\frac{s_{13}}{s_{12}}
+\frac{3}{2}\log\frac{s_{13}}{s_{12}}
\right.\right. \\ \nonumber
&&\left.~~~~~~~~~~~~~~~~~~
-\frac{1}{2}\frac{s_{13}}{s_{14}}\left[\left(1+\frac{s_{13}}{s_{14}}
\log\frac{s_{13}}{s_{12}}\right)^2
-\log\frac{s_{13}}{s_{12}}
+\left(\frac{s_{13}}{s_{14}}\right)^2\pi^2\right]\right] \\ \nonumber
&&~~~~~~~~~~~~-\frac{1}{N_c}\left[-\frac{1}{\varepsilon^2}
-\frac{3}{2\varepsilon}-4\right. \\ \nonumber
&&\left.\left.~~~~~~~~~~~~~~~~~~
-\frac{1}{2}\frac{s_{12}}{s_{14}}
\left[\left(1-\frac{s_{12}}{s_{14}}
\log\frac{s_{13}}{s_{12}}\right)^2
+\log\frac{s_{13}}{s_{12}}
+\left(\frac{s_{12}}{s_{14}}\right)^2\pi^2\right]\right]\right\}
-\frac{1}{\varepsilon}\beta_0.
\end{eqnarray}
\begin{equation}
c^{{\rm HV}}_{4;1}(+,-;+,-)=c_\Gamma
c_{4;0}(+,-;+,-)F^{+-}_{c;1}(\varepsilon, s_{12}, s_{13}, s_{14}),
\end{equation}
\begin{eqnarray}
\lefteqn{F^{+-}_{c;1}(\varepsilon,s_{12},s_{13},s_{14})=}\\ \nonumber
&&\left(-\frac{\mu^2}{s_{12}}\right)^\varepsilon
\left\{N_c\left[-\frac{3}{\varepsilon^2}-\frac{3}{2\varepsilon}
+\frac{2}{\varepsilon}\log\frac{s_{13}}{s_{12}}
-3+\pi^2+\frac{1}{2}\frac{s_{12}}{s_{14}}
\left(\log^2\frac{s_{13}}{s_{12}}+\pi^2\right)
\right]\right. \\ \nonumber
&&~~~~~~~~~~~~\left.-\frac{1}{N_c}\left[-\frac{1}{\varepsilon^2}
-\frac{3}{2\varepsilon}-4-\frac{1}{2}\frac{s_{12}}{s_{14}}
\left(\log^2\frac{s_{13}}{s_{12}}+\pi^2\right)
\right]\right\}-\frac{1}{\varepsilon}\beta_0.
\end{eqnarray}
\vspace*{10pt}
\begin{equation}
c^{{\rm HV}}_{4;2}(-,+;+,-)=c_\Gamma(-\i)\frac{\l 14\r}{[14]}[23][24]
F^{-+}_{c;2}(\varepsilon, s_{12}, s_{13}, s_{14}),
\end{equation}
\begin{eqnarray}
\lefteqn{F^{-+}_{c;2}(\varepsilon,s_{12},s_{13},s_{14})=}\\ \nonumber
&&-\frac{2}{\varepsilon}\left[\frac{1}{s_{12}}
\left(-\frac{\mu^2}{s_{12}}\right)^\varepsilon
\log\frac{s_{14}}{s_{13}}
+\frac{1}{s_{13}}
\left(-\frac{\mu^2}{s_{13}}\right)^\varepsilon
\log\frac{s_{14}}{s_{12}}\right]
+\frac{3}{2s_{14}}
\left[\log^2\frac{s_{13}}{s_{12}}+\pi^2\right].
\end{eqnarray}
\vspace*{10pt}
\begin{equation}
c^{{\rm HV}}_{4;2}(+,-;+,-)=c_\Gamma\i\frac{\l 24\r}{[24]}[13][14]
F^{+-}_{c;2}(\varepsilon, s_{12}, s_{13}, s_{14}),
\end{equation}
\begin{eqnarray}
\lefteqn{F^{+-}_{c;2}(\varepsilon,s_{12},s_{13},s_{14})=}\\ \nonumber
&&-\frac{2}{\varepsilon}\left[\frac{1}{s_{12}}
\left(-\frac{\mu^2}{s_{12}}\right)^\varepsilon
\log\frac{s_{14}}{s_{13}}
+\frac{1}{s_{14}}
\left(-\frac{\mu^2}{s_{14}}\right)^\varepsilon
\log\frac{s_{12}}{s_{13}}\right]
-\frac{3}{2s_{13}}
\left[\log^2\frac{s_{14}}{s_{12}}+\pi^2\right].
\end{eqnarray}
\vspace*{10pt}

\newpage
Process D:

\bigskip
\begin{equation}
d^{{\rm HV}}_{4;1}(-,-,+,+)=c_\Gamma
d_{4;0}(-,-,+,+)F^{--}_{d;1}(\varepsilon, s_{12}, s_{13}, s_{14}),
\end{equation}
\begin{eqnarray}
\lefteqn{F^{--}_{d;1}(\varepsilon,s_{12},s_{13},s_{14})=}\\ \nonumber
&&\left(-\frac{\mu^2}{s_{14}}\right)^\varepsilon
\left\{N_c\left[-\frac{4}{\varepsilon^2}-\frac{11}{3\varepsilon}
-\frac{2}{\varepsilon}\log\frac{s_{14}}{s_{12}}
-\frac{67}{9}+\pi^2\right]
+N_f\left[\frac{2}{3\varepsilon}+\frac{10}{9}\right]\right\}
-\frac{1}{\varepsilon}\beta_0.
\end{eqnarray}
\vspace*{10pt}
\begin{equation}
d^{{\rm HV}}_{4;1}(-,+,-,+)=c_\Gamma
d_{4;0}(-,+,-,+)F^{-+}_{d;1}(\varepsilon, s_{12}, s_{13}, s_{14}),
\end{equation}
\begin{eqnarray}
\lefteqn{F^{-+}_{d;1}(\varepsilon,s_{12},s_{13},s_{14})=}\\ \nonumber
&&\left(-\frac{\mu^2}{s_{14}}\right)^\varepsilon
\left\{N_c\left[-\frac{4}{\varepsilon^2}-\frac{11}{3\varepsilon}
-\frac{2}{\varepsilon}\log\frac{s_{14}}{s_{12}}
-\frac{67}{9}+\pi^2\right]
+N_f\left[\frac{2}{3\varepsilon}+\frac{10}{9}\right]\right.\\ \nonumber
&&~~~~~~~~~~~~-\left(N_c-N_f\right)\frac{s_{12}s_{14}}{s_{13}^2}
\left[1-\left(\frac{s_{12}}{s_{13}}-\frac{s_{14}}{s_{13}}\right)
\log\frac{s_{14}}{s_{12}}\right. \\ \nonumber
&&~~~~~~~~~~~~~~~~~~~~~~~~~~~~~~~~~~~~~~~~~
\left.-\left(\frac{s_{12}s_{14}}{s_{13}^2}-2\right)
\left(\log^2\frac{s_{14}}{s_{12}}+\pi^2\right)\right] \\ \nonumber
&&~~~~~~~~~~~~\left.+\left(\frac{11}{3}N_c-\frac{2}{3}N_f\right)
\frac{s_{14}}{s_{13}}\log\frac{s_{14}}{s_{12}}
-\frac{3}{2}N_f\frac{s_{12}s_{14}}{s_{13}^2}
\left(\log^2\frac{s_{14}}{s_{12}}+\pi^2\right)\right\}
-\frac{1}{\varepsilon}\beta_0.
\end{eqnarray}
\vspace*{10pt}
In these equations
\begin{equation}
c_\Gamma=\frac{1}{(4\pi)^{2-\varepsilon}}
\frac{\Gamma^2(1-\varepsilon)\Gamma(1+\varepsilon)}
{\Gamma(1-2\varepsilon)}
\end{equation}
is a ubiquitous prefactor, $V=N_c^2-1$.
For the sake of completeness, we list the tree level expressions here:
\begin{eqnarray}
a_{4;0}(-,-;+,+)&=&\i \frac{\l 12 \r [34]}{s_{14}}.\\
\vspaceinarray
a_{4;0}(-,+;-,+)&=&\i \frac{\l 13 \r [24]}{s_{14}}.\\
\vspaceinarray
\label{treegluonquark}
c_{4;0}(g,g;q^+,\bar{q}^-)&=&
-\i \frac{\l Iq\r\l I\bar{q}\r^3}
{\l 12 \r\l 2\bar{q} \r\l \bar{q}q \r\l q1 \r}.\\
\vspaceinarray
\label{treegluon}
d_{4;0}(g,g,g,g)&=&
\i\frac{\l IJ\r^4}{\l 12 \r\l 23 \r\l 34 \r\l 41 \r}.
\end{eqnarray}
For process C, $c_{4;0}$ is vanishing if the gluons have the same
helicity.  In eqs.\ (\ref{treegluonquark}) $I$ denotes the negative
helicity gluon. For process D, $d_{4;0}$ is non-vanishing only if two
gluons have positive and two of them have negative helicity. In eq.\
 (\ref{treegluon}), $I$ and $J$ denote the negative helicity gluons.

\subsection{Results obtained using dimensional reduction}

The one-loop color subamplitudes obtained using \DR can most
conveniently be written with the help of the results given in the
previous subsection. Obviously, to ${\cal O}(\varepsilon)$, only
the divergent amplitudes may differ in the various schemes, therefore,
we do not repeat the finite ones.

The difference in the algebra when \HV or \DR schemes are used
together with the tensor reduction of sect.\ 4, can be described by a
single parameter which is chosen to be $d$-dimensional in the \HV
scheme and has four-dimensional value in dimensional reduction.
Changing the value of this parameter, one finds the following
differences:
\begin{eqnarray}
a_{4;1}^{{\rm DR}}(-,\pm;\mp,+)-a_{4;1}^{{\rm HV}}(-,\pm;\mp,+)&=&
c_\Gamma a_{4;0}(-,\pm;\mp,+)
\left(\frac{2}{3}N_c-\frac{1}{N_c}\right).\\
\vspaceinarray
c_{4;1}^{{\rm DR}}(-,\pm;\mp,+)-c_{4;1}^{{\rm HV}}(-,\pm;\mp,+)&=&
c_\Gamma c_{4;0}(-,\pm;\mp,+)
\frac{1}{2}\left(N_c-\frac{1}{N_c}\right).\\
\vspaceinarray
d_{4;1}^{{\rm DR}}(-,\pm,\mp,+)-d_{4;1}^{{\rm HV}}(-,\pm,\mp,+)&=&
c_\Gamma d_{4;0}(-,\pm,\mp,+)\frac{1}{3}N_c,
\end{eqnarray}
while the $m_{4;2}$, ($m=a$, $c$ or $d$) subamplitudes are identical
in the two schemes. We remark that our results for the gluon-gluon
scattering without the fermionic contribution, obtained using
dimensional reduction is identical with the corresponding expression
obtained using string-based rules and the \FDH scheme in ref.\
\cite{Ber92}.

These differences may have two origins. They can arise from the
different regularization of both the \UV and the \IR singularities.
The difference in the \UV regularization can be absorbed into the
change of \lQCD. This difference can be understood as a finite
renormalization of the gauge coupling and can be calculated separately.
In the background
field technique we obtain the finite renormalization from calculating
the gluon self energy in two different regularization schemes and use
the background field Ward identity \cite{Abb81},
\begin{equation}
Z^f_g=(Z^f_A)^{-1/2}.
\end{equation}
The self energy has the well-known form,
\begin{equation}
\Pi^{ab}_{\mu\nu}(k)=
\i \delta^{ab}(k_\mu k_\nu-k^2g_{\mu\nu})\Pi(k^2),
\end{equation}
where
\begin{equation}
\Pi^{{\rm HV}}(k^2)=\left(\frac{g_r}{4\pi}\right)^2
\left(-\frac{4\pi\mu^2}{k^2}\right)^\varepsilon\Gamma(1+\varepsilon)
\left(-\frac{1}{\varepsilon}\beta_0-\frac{67N_c-10N_f}{9}\right)
+Z^f_A-1+{\cal O}(\varepsilon)
\end{equation}
in the \HV scheme and
\begin{equation}
\Pi^{{\rm DR}}(k^2)=\left(\frac{g_r}{4\pi}\right)^2
\left(-\frac{4\pi\mu^2}{k^2}\right)^\varepsilon\Gamma(1+\varepsilon)
\left(-\frac{1}{\varepsilon}\beta_0-\frac{64N_c-10N_f}{9}\right)
+{\cal O}(\varepsilon),
\end{equation}
in dimensional reduction. As a result of the finite renormalization, the
difference between the two results vanish, therefore,
\begin{equation}
Z^f_A=1+\frac{1}{3}N_c\left(\frac{g_r}{4\pi}\right)^2
\end{equation}
and
\begin{equation}
\label{finiteZgDR}
Z^f_g=1-\frac{1}{6}N_c\left(\frac{g_r}{4\pi}\right)^2.
\end{equation}
After this renormalization, the difference between the two schemes for
the amplitudes becomes
\begin{eqnarray}
\label{aDRminusHV}
\tilde{a}_{4;1}^{{\rm DR}}(-,\pm;\mp,+)-
a_{4;1}^{{\rm HV}}(-,\pm;\mp,+)&=&
c_\Gamma a_{4;0}(-,\pm;\mp,+)\left(N_c-\frac{1}{N_c}\right),\\
\vspaceinarray
\tilde{c}_{4;1}^{{\rm DR}}(-,\pm;\mp,+)-
c_{4;1}^{{\rm HV}}(-,\pm;\mp,+)&=&
c_\Gamma c_{4;0}(-,\pm;\mp,+)
\left(\frac{5}{6}N_c-\frac{1}{2N_c}\right),\\
\vspaceinarray
\label{dDRminusHV}
\tilde{d}_{4;1}^{{\rm DR}}(-,\pm,\mp,+)-
d_{4;1}^{{\rm HV}}(-,\pm,\mp,+)&=&
c_\Gamma d_{4;0}(-,\pm,\mp,+)\frac{2}{3}N_c.
\end{eqnarray}
The tilde in eqs.\ (\ref{aDRminusHV}--\ref{dDRminusHV}) reminds
us that the expansion parameter is the same $\alpha_{\overline{{\rm
MS}}}$ in both schemes. In sect.\ 10, we
shall discuss how these differences can be incorporated into the
definition of the parton densities.

As a final step in this section, following ref.\ \cite{Cel79}, we
calculate the change in \lQCD\ caused by the finite renormalization
of the gauge coupling according to eq.\ (\ref{finiteZgDR}):
\begin{equation}
\Lambda_{{\rm DR}}=
\Lambda_{\overline{\rm MS}}\exp\frac{c}{\beta_0}.
\end{equation}
The constant $c$ is defined by
\begin{equation}
\alpha_s^{{\rm DR}}=\alpha_s^{\overline{\rm MS}}
\left(1+c\frac{\alpha_s^{\overline{\rm MS}}}{2\pi}\right),
\end{equation}
i.e., $c=N_c/6$ in agreement with the result of ref.\
\cite{Alt81} obtained in Lorentz-Feynman gauge.

\section{Loop contribution to the \NLO\ matrix element}
\setcounter{equation}{0}

In this section, first we give explicit formul\ae\
how to obtain the loop contribution to the \NLO\ matrix element
summed over color and helicities and
then compare the obtained results with the corresponding
expressions given in ref.\ \cite{Ell86}. In this way, we can
establish the transition rules for loop amplitudes which connect
our results to those obtained in the \CDR scheme.

\subsection{Squared matrix element in the \HV scheme}

It is useful to rewrite the $F$ functions in terms
of the physical variables $s$, $t$ and $u$ in a form which will be more
convenient when the square of the matrix element is calculated.
In the \NLO\ amplitudes of sect.\ 5, we factored out the Born terms.
Therefore, it is sufficient to present only the real part of the $F$
functions since the imaginary part will not play any role in a \NLO\
calculation of the squared matrix elements (${\cal R}e F\equiv \F$).
To keep the crossing symmetry of the matrix element manifest, we leave
the sign of $s$, $t$ and $u$ open, and introduce an auxiliary mass
variable $Q$ to express the logarithms \cite{Ell86}. As mentioned
previously, the analytic continuation is done according to formula
(\ref{continue}). To keep the real parts of the logarithms, we use
the following substitutions:
\begin{equation}
\log\left(\frac{s_{12}}{s_{14}}\right)\ra\ell(s)-\ell(t),\qquad
\log^2\left(\frac{s_{12}}{s_{14}}\right)\ra\ell_2(s)+\ell_2(t) -
2\ell(s)\ell(t),
\end{equation}
\begin{equation}
\log\left(\frac{s_{12}}{s_{13}}\right)\ra\ell(s)-\ell(u),\qquad
\log^2\left(\frac{s_{12}}{s_{13}}\right)\ra\ell_2(s)+\ell_2(u) -
2\ell(s)\ell(u),
\end{equation}
\begin{equation}
\log\left(\frac{s_{13}}{s_{14}}\right)\ra\ell(u)-\ell(t),\qquad
\log^2\left(\frac{s_{13}}{s_{14}}\right)\ra\ell_2(u)+\ell_2(t) -
2\ell(u)\ell(t),
\end{equation}
where
\begin{equation}
\ell(x)=\log\left|\frac{x}{Q^2}\right|\quad{\rm and}\quad
\ell_2(x)=\ell^2(x)-\pi^2\Theta(x).
\end{equation}
One finds the following results.

\bigskip
Process A:

\bigskip
For the square of the matrix element only
the sum of the $\F_{a;1}$ and $\F_{a;2}$ functions is needed.
\begin{eqnarray}
\lefteqn{\F^{--}_a(\varepsilon,s,t,u)\equiv
{\cal R}e[F^{--}_{a;1}(\varepsilon,s,t,u)
+F^{--}_{a;2}(\varepsilon,s,t,u)]=} \\ \nonumber
&&\left(\frac{\mu^2}{Q^2}\right)^\varepsilon
\left\{\frac{V}{N_c}
\left(-\frac{2}{\varepsilon^2}-\frac{3}{\varepsilon}\right)-
\frac{10}{9}N_f+\frac{V}{N_c}\left(
\frac{2}{\varepsilon}\ell(t)-\ell_2(t)+3\ell(t)\right)+
\frac{2}{3}N_f\ell(t)\right.\\ \nonumber
&&~~~~~~~~~~+N_c\left[\frac{2}{\varepsilon}(\ell(u)-\ell(t))+
\frac{13}{9}+\pi^2
+2\ell_2(t)-2\ell(t)\ell(u)-\frac{11}{3}\ell(t)\right. \\ \nonumber
&&~~~~~~~~~~~~~~~~\left.+\frac{t}{2s}\left(1-\frac{u}{s}\right)
\left(\ell_2(t)+\ell_2(u)+\pi^2-2\ell(t)\ell(u)\right)
+\frac{t}{s}(\ell(t)-\ell(u))\right] \\ \nonumber
&&~~~~~~~~~~+\frac{1}{N_c}\left[\frac{4}{\varepsilon}(\ell(s)-\ell(u))+
8+2\frac{t}{s}(\ell(u)-\ell(t))\right. \\ \nonumber
&&~~~~~~~~~~~~~~~~\left.\left.-\frac{t}{s}\left(1-\frac{u}{s}\right)
\left(\ell_2(t)+\ell_2(u)+\pi^2-2\ell(t)\ell(u)\right)
-4\ell(s)\ell(t)+4\ell(t)\ell(u)\right]\right\} \\ \nonumber
&&~~~~~~~~~~
-\frac{1}{\varepsilon}\beta_0.
\end{eqnarray}
\begin{eqnarray}
\lefteqn{\F^{-+}_a(\varepsilon,s,t,u)\equiv
{\cal R}e[F^{+-}_{a;1}(\varepsilon,s,t,u)
+F^{-+}_{a;2}(\varepsilon,s,t,u)]=} \\ \nonumber
&&\left(\frac{\mu^2}{Q^2}\right)^\varepsilon
\left\{\frac{V}{N_c}
\left(-\frac{2}{\varepsilon^2}-\frac{3}{\varepsilon}\right)-
\frac{10}{9}N_f+\frac{V}{N_c}\left(
\frac{2}{\varepsilon}\ell(t)-\ell_2(t)+3\ell(t)\right)+
\frac{2}{3}N_f\ell(t)\right.\\ \nonumber
&&~~~~~~~~~~+N_c\left[\frac{2}{\varepsilon}(\ell(u)-\ell(t))+
\frac{13}{9}+\pi^2
+2\ell_2(t)-2\ell(t)\ell(u)-\frac{11}{3}\ell(t)\right] \\ \nonumber
&&~~~~~~~~~~+\frac{1}{N_c}\left[\frac{4}{\varepsilon}(\ell(s)-\ell(u))+
8-4\ell(s)\ell(t)+4\ell(t)\ell(u)\right. \\ \nonumber
&&~~~~~~~~~~~~~~~~\left.\left.+\frac{t}{u}\left(1-\frac{s}{u}\right)
\left(\ell_2(t)+\ell_2(s)+\pi^2-2\ell(t)\ell(s)\right)
+2\frac{t}{u}(\ell(t)-\ell(s))\right]\right\} \\ \nonumber
&&~~~~~~~~~~
-\frac{1}{\varepsilon}\beta_0.
\end{eqnarray}

\bigskip
Process B:

\bigskip
For the square of the matrix element, we shall need $\F^{--}_{a;1}$
in addition to the $\F_a$ functions.
\begin{eqnarray}
\lefteqn{\F^{--}_{a;1}(\varepsilon,s,t,u)=} \\ \nonumber
&&\left(\frac{\mu^2}{Q^2}\right)^\varepsilon
\left\{\frac{V}{N_c}
\left(-\frac{2}{\varepsilon^2}-\frac{3}{\varepsilon}\right)-
\frac{10}{9}N_f+\frac{V}{N_c}\left(
\frac{2}{\varepsilon}\ell(t)-\ell_2(t)+3\ell(t)\right)+
\frac{2}{3}N_f\ell(t)\right.\\ \nonumber
&&~~~~~~~~~~+N_c\left[\frac{2}{\varepsilon}(\ell(s)-\ell(t))+
\frac{13}{9}+\pi^2
+2\ell_2(t)-2\ell(t)\ell(s)-\frac{11}{3}\ell(t)\right] \\ \nonumber
&&~~~~~~~~~~+\frac{1}{N_c}\left[\frac{2}{\varepsilon}(\ell(s)-\ell(u))+
8-2\ell(s)\ell(t)+2\ell(t)\ell(u)\right. \\ \nonumber
&&~~~~~~~~~~~~~~~~\left.\left.-\frac{t}{2s}\left(1-\frac{u}{s}\right)
\left(\ell_2(t)+\ell_2(u)+\pi^2-2\ell(t)\ell(u)\right)
+\frac{t}{s}(\ell(u)-\ell(t))\right]\right\} \\ \nonumber
&&~~~~~~~~~~
-\frac{1}{\varepsilon}\beta_0.
\end{eqnarray}

\bigskip
Process C:

\bigskip
\begin{eqnarray}
\lefteqn{\F^{+-}_{c;1}(\varepsilon,s,t,u)=
\left(\frac{\mu^2}{Q^2}\right)^\varepsilon
\left\{\frac{V}{N_c}
\left(-\frac{1}{\varepsilon^2}-\frac{3}{2\varepsilon}
-\frac{7}{2}\right)-N_c\frac{2}{\varepsilon^2}\right.} \\ \nonumber
&&~~~~~~~~~~~~~~~~~~~~~~~~~~+\frac{V}{N_c}
\left(\frac{1}{\varepsilon}\ell(s)+\frac{3}{2}\ell(s)
-\frac{1}{2}\ell_2(s)\right)
+N_c\left(\frac{2}{\varepsilon}\ell(u)
          +\ell_2(s)+\pi^2\right) \\ \nonumber
&&~~~~~~~~~~~~~~~~~~~~~~~~~~\left.-2N_c\ell(s)\ell(u)
+\frac{N_c^2+1}{2N_c}
\left(1+\frac{s}{t}\left(\ell_2(s)+\ell_2(u)-2\ell(s)\ell(u)+\pi^2
\right)\right)\right\} \\ \nonumber
&&~~~~~~~~~~~~~~~~~~-\frac{1}{\varepsilon}\beta_0.
\end{eqnarray}
\begin{eqnarray}
\lefteqn{\F^{-+}_{c;1}(\varepsilon,s,t,u)=} \\ \nonumber
&&\left(\frac{\mu^2}{Q^2}\right)^\varepsilon
\left\{\frac{V}{N_c}
\left(-\frac{1}{\varepsilon^2}-\frac{3}{2\varepsilon}
-\frac{7}{2}\right)-N_c\frac{2}{\varepsilon^2}~~~~~~~~~~\right. \\ \nonumber
&&~~~~~~~~~~+\frac{V}{N_c}
\left(\frac{1}{\varepsilon}\ell(s)+\frac{3}{2}\ell(s)
-\frac{1}{2}\ell_2(s)\right)
+N_c\left(\frac{2}{\varepsilon}\ell(u)+\ell_2(s)+\pi^2\right) \\ \nonumber
&&~~~~~~~~~~\left.-\frac{1}{2}N_c
\right[\ell_2(s)+\ell_2(u)+\pi^2+2\ell(s)\ell(u)
      +3\ell(s)-3\ell(u) \\ \nonumber
&&~~~~~~~~~~~~~~~~~~~~~~~~+\frac{u}{t}
\left(1+\left(1-2\frac{u}{t}\right)\left(\ell(s)-\ell(u)\right)\right.\\
\nonumber &&~~~~~~~~~~~~~~~~~~~~~~~~~~~\left.\left.
+\left(\frac{u}{t}\right)^2
\left(\ell_2(s)+\ell_2(u)-2\ell(s)\ell(u)+\pi^2\right)\right)
\right] \\ \nonumber
&&~~~~~~~~~~~+\frac{1}{2N_c}
\left[1+\frac{s}{t}
\left(1-\left(1-2\frac{s}{t}\right)
\left(\ell(s)-\ell(u)\right)\right.\right. \\
\nonumber &&~~~~~~~~~~~~~~~~~~~~~~~~~~~\left.\left.\left.
+\left(\frac{s}{t}\right)^2
\left(\ell_2(s)+\ell_2(u)-2\ell(s)\ell(u)+\pi^2\right)\right)
\right]\right\} \\ \nonumber
&&-\frac{1}{\varepsilon}\beta_0.
\end{eqnarray}
It turns out that only the sum and difference of the $\F_{c;2}$
functions are needed.
\begin{eqnarray}
\lefteqn{\F^+_c(\varepsilon,s,t,u)\equiv
{\cal R}e(F^{-+}_{c;2}(\varepsilon,s,t,u)
         +F^{+-}_{c;2}(\varepsilon,s,t,u))=}\\ \nonumber
&&\left(\frac{\mu^2}{Q^2}\right)^\varepsilon
\left\{\frac{2}{\varepsilon}\left[
\frac{2}{s}\left(\ell(u)-\ell(t)\right)
+\frac{1}{t}\left(\ell(u)-\ell(s)\right)
+\frac{1}{u}\left(\ell(s)-\ell(t)\right)\right]\right. \\ \nonumber
&&~~~~~~~~~~\left.\left.
+\frac{4}{s}\right[\ell(s)\ell(t)-\ell(s)\ell(u)\right] \\
\nonumber &&~~~~~~~~~~+\frac{1}{t}\left[
\frac{3}{2}\left(\ell_2(s)+\ell_2(u)+\pi^2\right)-3\ell(s)\ell(u)
                 +2\ell(s)\ell(t)-2\ell(t)\ell(u)\right] \\
\nonumber &&~~~~~~~~~~\left.-\frac{1}{u}\left[
\frac{3}{2}\left(\ell_2(s)+\ell_2(t)+\pi^2\right)-3\ell(s)\ell(t)
                 +2\ell(s)\ell(u)-2\ell(t)\ell(u)\right]\right\},
\end{eqnarray}
\begin{eqnarray}
\lefteqn{\F^-_c(\varepsilon,s,t,u)\equiv
{\cal R}e(F^{-+}_{c;2}(\varepsilon,s,t,u)
         -F^{+-}_{c;2}(\varepsilon,s,t,u))=}\\ \nonumber
&&\left(\frac{\mu^2}{Q^2}\right)^\varepsilon
\left\{\frac{2}{\varepsilon}\left[
\frac{1}{t}\left(\ell(s)-\ell(u)\right)
+\frac{1}{u}\left(\ell(s)-\ell(t)\right)\right]\right. \\ \nonumber
&&~~~~~~~~~~+\frac{1}{t}\left[
\frac{3}{2}\left(\ell_2(s)+\ell_2(u)+\pi^2\right)-3\ell(s)\ell(u)-
                 2\ell(s)\ell(t)+2\ell(t)\ell(u)\right] \\
\nonumber &&~~~~~~~~~~\left.+\frac{1}{u}\left[
\frac{3}{2}\left(\ell_2(s)+\ell_2(t)+\pi^2\right)-3\ell(s)\ell(t)-
                 2\ell(s)\ell(u)+2\ell(t)\ell(u)\right]\right\}.
\end{eqnarray}

\bigskip
Process D:

\bigskip
\begin{eqnarray}
\lefteqn{\F^{--}_{d;1}(\varepsilon,s,t,u)=}\\ \nonumber
&&\left(\frac{\mu^2}{Q^2}\right)^\varepsilon
\left\{N_c\left[-\frac{4}{\varepsilon^2}-\frac{11}{3\varepsilon}
+\frac{2}{\varepsilon}\left(\ell(s)+\ell(t)\right)
-\frac{67}{9}+\pi^2+\frac{11}{3}\ell(t)
-2\ell(s)\ell(t)\right]\right. \\ \nonumber
&&~~~~~~~~~~\left.
+N_f\left[\frac{2}{3\varepsilon}+\frac{10}{9}
-\frac{2}{3}\ell(t)\right]\right\}
-\frac{1}{\varepsilon}\beta_0.
\end{eqnarray}
\vspace*{10pt}
\begin{eqnarray}
\lefteqn{\F^{-+}_{d;1}(\varepsilon,s,t,u)=}\\ \nonumber
&&\left(\frac{\mu^2}{Q^2}\right)^\varepsilon
\left\{N_c\left[-\frac{4}{\varepsilon^2}-\frac{11}{3\varepsilon}
+\frac{2}{\varepsilon}\left(\ell(s)+\ell(t)\right)
-\frac{67}{9}+\pi^2+\frac{11}{3}\ell(t)
-2\ell(s)\ell(t)\right]\right. \\ \nonumber
&&~~~~~~~~~~
+N_f\left[\frac{2}{3\varepsilon}+\frac{10}{9}
-\frac{2}{3}\ell(t)\right]\\ \nonumber
&&~~~~~~~~~~-\left(N_c-N_f\right)\frac{st}{u^2}
\left[1-\left(\frac{s}{u}-\frac{t}{u}\right)
\left(\ell(t)-\ell(s)\right)\right. \\ \nonumber
&&~~~~~~~~~~~~~~~~~~~~~~~~~~\left.-\left(\frac{st}{u^2}-2\right)
\left(\ell_2(s)+\ell_2(t)+\pi^2-2\ell(s)\ell(t)\right)\right] \\ \nonumber
&&~~~~~~~~~~~~\left.+\beta_0
\frac{t}{u}\left(\ell(t)-\ell(s)\right)
-\frac{3}{2}N_f\frac{st}{u^2}
\left(\ell_2(s)+\ell_2(t)+\pi^2-2\ell(s)\ell(t)\right)\right\} \\ \nonumber
&&~~~~~~~~~~~~-\frac{1}{\varepsilon}
\beta_0.
\end{eqnarray}

The square of the matrix element summed over helicities can be
expressed in terms of the $\F$ functions.

\bigskip
Process A:

\bigskip
\begin{equation}
\sum_{\rm hel}\sum_{\rm col} [\M^*\M]_{\rm NLO}
= 2g^6 c_\Gamma 2V
\left(\frac{s^2}{t^2}\F^{--}_a(s,t,u)
+\frac{u^2}{t^2}\F^{-+}_a(s,t,u)\right).
\end{equation}

\bigskip
Process B:

\bigskip
\begin{eqnarray}
\lefteqn{\sum_{\rm hel}\sum_{\rm col} [\M^*\M]_{\rm NLO}
= 2g^6 c_\Gamma 2V} \\ \nonumber
&&\left(\frac{s^2}{t^2}\F^{--}_a(s,t,u)
+\frac{u^2}{t^2}\F^{-+}_a(s,t,u)
+\frac{s^2}{u^2}\F^{--}_a(s,u,t)
+\frac{t^2}{u^2}\F^{-+}_a(s,u,t)\right. \\ \nonumber
&&~~~~~~~~~~~~~~~~~~~~\left.
-\frac{1}{N_c}\frac{s^2}{tu}\left(\F^{--}_{a;1}(s,t,u)
+\F^{--}_{a;1}(s,u,t)\right)\right).
\end{eqnarray}

\bigskip
Process C:

\bigskip

\begin{eqnarray}
\lefteqn{\sum_{\rm hel}\sum_{\rm col} [\M^*\M]_{\rm NLO}
= 2g^6 c_\Gamma \frac{V}{N_c}} \\ \nonumber
&&\times\left\{2V\frac{1}{s^2}\left[
tu\left(\F^{+-}_{c;1}(s,t,u)+\F^{+-}_{c;1}(s,u,t)\right)
+\frac{t^3}{u}\F^{-+}_{c;1}(s,t,u)
+\frac{u^3}{t}\F^{-+}_{c;1}(s,u,t)\right]\right. \\ \nonumber
&&~~~~~~~~~~~~~~~-\left.\left.\frac{2}{s^2}\right[
t^2\left(\F^{-+}_{c;1}(s,t,u)+\F^{+-}_{c;1}(s,u,t)\right)
+u^2\left(\F^{+-}_{c;1}(s,t,u)+\F^{-+}_{c;1}(s,u,t)\right)\right]\\ \nonumber
&&~~~~~~~~~~~~~-\left.\left.\left.N_c\frac{1}{s}
\right[s^2 \F^-_c(s,t,u)+t^2 \F^+_c(s,t,u)-u^2 \F^+_c(s,t,u)\right]\right\}.
\end{eqnarray}

\bigskip
Process D:

\bigskip
\begin{equation}
\sum_{\rm hel}\sum_{\rm col} [\M^*\M]_{\rm NLO}
= 2g^6 c_\Gamma 2N_c^2V
\left(\F_d(s,t,u)+\F_d(t,u,s)+\F_d(u,s,t)\right),
\end{equation}
where
\begin{equation}
\F_d(s,t,u)=
2\frac{s^2}{t^2}\F^{--}_{d;1}(s,t,u)
+2\frac{s^2}{u^2}\F^{--}_{d;1}(s,u,t)
+\frac{s^4}{t^2u^2}
\left(\F^{-+}_{d;1}(u,t,s)+\F^{-+}_{d;1}(t,u,s)\right).
\end{equation}

\subsection{Comparison with existing results}

The matrix elements given in the previous subsection differ in finite
terms from those calculated in the conventional regularization scheme
by Ellis and Sexton
\cite{Ell86}. The difference can easily be described if one rewrites
the Ellis-Sexton matrix elements in the form given in ref.\ \cite{Kun92}.

Following ref.\ \cite{Kun92}, we denote a certain cross section
(e.g., inclusive one-jet cross section) by $\I$. At \NLO, $\I$ is a
sum of two terms,
\begin{equation}
\I=\I[2\ra2]+\I[2\ra3],
\end{equation}
where $\I[2\ra n]$ is the $[2\ra n]$ part of the cross section.
According to the factorization theorem, the physical cross section in
the QCD improved parton model for hadron-hadron scattering is a
folding between the parton densities and the hard-scattering cross
section:
\begin{equation}
\label{factorization}
\I[2\ra n]=\sum_{a,b}
\int_0^1\d x_a\int_0^1\d x_b f_{a/A}(x_a,\mu)f_{b/B}(x_b,\mu)
\d\hat{\sigma}_{a,b}(x_a p_A, x_b p_B, \mu, \alpha_s(\mu)).
\end{equation}
In eq.\ (\ref{factorization}), $\d\hat{\sigma}_{a,b}$ is the
hard-scattering cross section for the process $a+b\ra j_1+\ldots+j_n$.
It is defined as a product of
the flux factor and the integral of the squared matrix element
over the phase space of the final state particles
\begin{eqnarray}
\lefteqn{\d\hat{\sigma}_{a,b}=
\frac{1}{n!}\sum_{j_1,\ldots,j_n}\frac{1}{2x_ax_bs}} \\ \nonumber
&&\int\d PS^{(n)}({\rm {\bf p}}_{j_i})\,
\S_n(p_{j_i}^\mu)\l|\M(a+b\ra j_1+\ldots j_n)|^2\r
(2\pi)^d\delta^d\left(p_a^\mu+p_b^\mu-\sum_{i=1}^n p_{j_i}^\mu\right),
\end{eqnarray}
where $\S_n(p_{j_i}^\mu)$ is the so called measurement function that
defines the infrared-safe physical quantity. The counting factors $1/n!$
are present when all partons are treated indistinguishable and we sum
over the possible parton types.  For the $[2\ra2]$ process,
the square of the matrix element --- summed over final spins and
colors and averaged over initial spins and colors --- has the
following perturbative expansion
\begin{equation}
\l|\M(a+b\ra j_1+j_2|^2\r=\frac{g^4}{\omega(a)\omega(b)}
\left\{\psi^{(4)}(\vec{a},\vec{p})
+2g^2\left(\frac{\mu^2}{Q^2}\right)^{\varepsilon}
c_\Gamma\psi^{(6)}(\vec{a},\vec{p})
+{\cal O}(g^4)\right\},
\end{equation}
where we denote $\vec{a}=(a,b,j_1,j_2)$, $\vec{p}=(p_a^\mu,
p_b^\mu,p_{j_1}^\mu,p_{j_2}^\mu)$ and $\omega(a)$
represents the number of spin and color states of a parton type
$a$. Comparing our notation to the one just defined, we see that
\begin{equation}
2g^6\left(\frac{\mu^2}{Q^2}\right)^{\varepsilon}
c_\Gamma\psi^{(6)}(\vec{a},\vec{p})\equiv
\sum_{{\rm hel}}\sum_{{\rm col}}[\M^*\M]_{{\rm NLO}}.
\end{equation}

In ref.\ \cite{Kun92}, using the results of Ellis and Sexton,
the following structure has been found for the \NLO\ term:
\begin{eqnarray}
\label{Kunsztform}
\psi^{(6)}(\vec{a},\vec{p})&=&
\psi^{(4)}(\vec{a},\vec{p})
\left\{-\frac{1}{\varepsilon^2}\sum_n C(a_n)
-\frac{1}{\varepsilon}\sum_n \gamma(a_n)\right\} \\ \nonumber
&&+\frac{1}{\varepsilon}\sum_{m<n}
\log\left(\frac{2 p_n\cdot p_m}{Q^2}\right)
\psi^{(4,c)}_{mn}(\vec{a},\vec{p}) \\ \nonumber
&&+\psi^{(6)}_{{\rm NS}}(\vec{a},\vec{p}),
\end{eqnarray}
where $\psi^{(4)}(\vec{a},\vec{p})$ is
the $d$-dimensional Born term, $\psi^{(4,c)}_{mn}
(\vec{a},\vec{p})$ are the color-linked Born squared matrix elements
in $d$ dimensions as defined in Appendix A of ref.\ \cite{Kun92} and
$\psi^{(6)}_{{\rm NS}}(\vec{a},\vec{p})$ represents the remaining
finite terms. The sum over $m$ and $n$ runs from one to four. In eq.\
(\ref{Kunsztform}), $C(a)$ is the color charge of
parton $a$ and the constant $\gamma(a)$ represents the contribution
from virtual diagrams to the \AP kernel. Specifically,
\begin{equation}
C(g)=N_c,\; \gamma(g)=\frac{1}{2}\beta_0,
\end{equation}
\begin{equation}
C(q)=\frac{V}{2N_c},\; \gamma(q)=\frac{3V}{4N_c}.
\end{equation}

Using this notation, the transition from the conventional scheme to the
't Hooft-Veltman scheme can be achieved by substituting the
four-dimensional expressions for the $\psi^{(4)}$ and
$\psi^{(4,c)}_{mn}$ functions and leaving the $\psi^{(6)}_{\rm NS}$
functions unchanged. In sect.\ 8, we shall analyze the
structure of the squared matrix element for the Bremsstrahlung part of
the cross section and shall point out that part of the cross section
undergoes analogous change when passing from the \CDR to the \HV
scheme.

Next, we consider the squared matrix element in the \HV and \DR schemes
in the form as given in eq.\ (\ref{Kunsztform}). The difference described
at amplitude level by eqs.\ (\ref{aDRminusHV}--\ref{dDRminusHV})
means a difference in the $\psi^{(6)}_{{\rm NS}}$
functions. We see, however, that the difference is proportional to the
Born term.  Conseqently, we can use the same (properly modified)
$\psi^{(6)}_{{\rm NS}}$ functions in both schemes and include the
difference into a scheme dependent modification of the $\gamma(a)$
functions. Instead of $\gamma(a)$, we use the following
$\varepsilon$-dependent $\gamma(a,\varepsilon)$ functions:
\begin{equation}
\gamma(a,\varepsilon)=
\gamma(a)+\varepsilon\tilde{\gamma}(a),
\end{equation}
where
\begin{equation}
\label{gammatilde}
\tilde{\gamma}(g)=\frac{1}{6}N_c,\;
\tilde{\gamma}(q)=\frac{1}{2}C_F,
\end{equation}
with $C_F=V/(2N_c)$. We obtained the above expressions for the
$\tilde{\gamma}(a)$ terms from the differences among the squared
matrix elements in the \HV and \DR schemes for all processes.
These differences determine an overconstrained system of linear
equations (four equations for two unknowns). We could solve
this overconstrained system consistently, which indicates that using
the $\gamma(a,\varepsilon)$ functions in eq.\ (\ref{Kunsztform}),
universality will be maintained. We receive further indication of
universality if we recall our observation that for the four-gluon
amplitudes in pure gauge theory, one obtains identical
results using the string-based rules and \FDH scheme as in field
theory using dimensional reduction. Extrapolating this rule to the
five-gluon amplitudes, one can easily check, using the results of
ref.\ \cite{Ber93}, that universality is maintained at the five-point
level. We have an additional confirmation that the
$\gamma(a,\varepsilon)$ functions are process independent. We can
obtain them if we extend the validity of the momentum sum-rule to the
$\varepsilon$-dependent part of the \AP kernels in the \HV scheme, as
can easily be checked using the expressions of Appendix B.

When we write the results for the squared matrix element in the form of
eq.\ (\ref{Kunsztform}), but in terms of $\gamma(a,\varepsilon)$
functions, then the transition from the \HV scheme to \DR is
accomplished by setting
$\tilde{\gamma}(a)=0$. This observation will be important when we
discuss how to shift the difference in the loop amplitudes described
by $\tilde{\gamma}(a)$ into the parton density functions.

\section{Scattering~ amplitudes~ in~ \N1~ \YM  theory at one loop}
\setcounter{equation}{0}

The classical Lagrangian of the \N1  \YM  theory in Wess-Zumino gauge
for the component fields reads \cite{Wes74}
\begin{equation}
{\cal L}=-\frac{1}{4}(F^a_{\mu\nu})^2
-\frac{1}{2}\bar{\lambda}^a\slash{{\cal D}}\lambda_a,
\end{equation}
where $F^a_{\mu\nu}$ is the usual \YM  field strength of a vector
field $g$ in the adjoint representation, ${\cal D}$ is the usual
covariant derivative and $\lambda$ is a Weyl spinor in the adjoint
representation. From this Lagrangian one immediately sees that the
kinematic structure of the gluon-gluino-gluino coupling in \N1 \YM
theory is the
same as the gluon-quark-antiquark coupling in QCD. The only difference
between these two couplings is the color charge: in QCD the quarks are
in the fundamental representation while in the supersymmetric theory
the gluinos are in the adjoint representation. Therefore, changing the
color matrices $T^a$ in our previous calculation for the QCD
amplitudes to $F^a$, we can obtain the gluino-gluino,
gluon-gluino and gluon-gluon scattering amplitudes in \N1
\YM  theory. Such a procedure has a direct application in tree-level QCD
calculation: the subamplitudes for gluon-gluon scattering can be
obtained from those of quark-gluon scattering
\cite{Par85,Kun86}. At one-loop level the color structure of the
amplitudes in QCD does not allow for such a direct use of
supersymmetry. One obvious application is going into the
supersymmetry limit and use \SWI\ as important checks on the
calculation (see below). A more direct application is the following.
Calculate the quark-gluon scattering amplitudes, change the color
charges to obtain the gluino-gluon scattering amplitudes. Using \SWI,
obtain the gluon-gluon scattering amplitudes in \N1 \YM theory.
Calculate that part of the gluon-gluon scattering amplitude in QCD which
is proportional to the number of quark flavors, $N_f$. This calculation
is much simpler than the complete calculation because it involves the
evaluation of only those Feynman diagrams which contain a closed fermion
loop. Subtract this contribution from the result in the supersymmetric
theory with color charges corresponding to the adjoint representation of
the fermions (i.e, $N_f\ra N_c$) and simultaneously add it with the QCD
color charges. Thus we obtain the correct amplitudes for gluon-gluon
scattering in QCD saving considerable amount of work in that part of the
calculation which is the most difficult to carry out.

In order to obtain a meaningful result in a supersymmetric theory
beyond tree level, one has to use a supersymmetric regulator or
alternatively one has to restore the \SWI\ if a supersymmetry
breaking regularization has been used. At one-loop level,
dimensional reduction is known to respect supersymmetry. It is,
however, interesting to see, how supersymmetry can be restored when a
non-supersymmetry-preserving regulator (such as the \HV scheme) is used.
Therefore, we shall elaborate on both approaches.

Let us now give the results of the calculation performed in the \DR
scheme. We shall use a self-evident extension of naming the processes,
i.e., process B will be the gluino-gluino scattering, process C will
be the gluon-gluino scattering and process D will be the gluon-gluon
scattering. The results will be given for one gluino flavor.
In a supersymmetric theory, the amplitudes for those helicity
configurations which are vanishing at tree level have to vanish at one
loop as well. Explicit calculation shows that this property is fulfilled in
the one-loop calculation. In the following, we give only those
non-vanishing amplitudes from which one can obtain other amplitudes
using parity transformation or cyclic property of the amplitudes.
As usual, all particles are outgoing. The results can be given in a
concise form:
\begin{equation}
\label{SUSYresultDR}
m^{{\rm SUSY,DR}}_{4;1}(-,\mp,\pm,+)=
c_\Gamma m^{{\rm SUSY}}_{4;0}(-,\mp,\pm,+)
N_cF^{-\mp}_1(\varepsilon, s_{12}, s_{13}, s_{14}),
\end{equation}
where $m$ stands for $b$, $c$ or $d$. The $F^{-\mp}_1$ functions
are universal:
\begin{eqnarray}
\lefteqn{F^{-\lambda}_1(\varepsilon,s_{12},s_{13},s_{14})=} \\ \nonumber
&&\left(-\frac{\mu^2}{s_{14}}\right)^\varepsilon
\left(-\frac{4}{\varepsilon^2}-\frac{3}{\varepsilon}
-\frac{2}{\varepsilon}\log\frac{s_{14}}{s_{12}}
-6+\pi^2\right)-\frac{3}{\varepsilon} \\ \nonumber
&&~~~~~~~~~~~+\delta^{+\lambda}
\left(3\frac{s_{14}}{s_{13}}\log\frac{s_{14}}{s_{12}}
-\frac{3}{2}\frac{s_{12}s_{14}}{s_{13}^2}
\left(\log^2\frac{s_{14}}{s_{12}}+\pi^2\right)\right).
\end{eqnarray}

It is easy to check that these results are indeed supersymmetric,
i.e., they satisfy certain on-shell \SWI.
These Ward identities can easily be derived noting that the
supersymmetry charge $Q(\eta)$ with $\eta$ being the fermionic
parameter of the transformation annihilates the vacuum. Then it
follows that the commutator of $Q(\eta)$ with any string of operators
creating or annihilating of a gluon or a gluino  has a vanishing
vacuum expectation value \cite{Gri77}. This statement is true in any
order of perturbation theory. If $a_i$ represent these operators,
then we find the following \SWI:
\begin{equation}
\label{SWI}
0=\left\l\left[Q,\prod_{i=1}^n a_i\right]\right\r_0=
\sum_{i=1}^n\l a_1\ldots [Q,a_i]\ldots a_n\r_0.
\end{equation}
In the \N1  \YM  theory we considered above, the $a_i$ stand for
$g^\pm$ and $\lambda^\pm$. The superscripts $\pm$ refer to the two
possible helicity states of the vector and spinor. The action of the
supersymmetry charge on the doublet ($g$, $\lambda$) is as follows
\cite{Gri77}:
\begin{equation}
[Q(\eta), g^\pm(p)]=\mp\Gamma^\pm(p,\eta)\lambda^\pm.
\end{equation}
\begin{equation}
[Q(\eta), \lambda^\pm(p)]=\mp\Gamma^\mp(p,\eta)g^\pm.
\end{equation}
Substituting these commutation relations into eq.\ (\ref{SWI}) we
obtain a relation among various scattering amplitudes for particles
with different spin. These relations are the on-shell \SWI\ referred
to above.

We remark that the \SWI\ hold separately for each of the
subamplitudes in which one can expand the full amplitude \cite{Man91}.

In order to find explicit relations, the $\Gamma^\pm(p,\eta)$
functions are to be specified yet. $\Gamma^\pm(p,\eta)$ is a complex
function linear in the anticommuting $c$-number components of $\eta$
and satisfies
\begin{equation}
\Gamma^+(p,\eta)=(\Gamma^-(p,\eta))^*=\bar{\eta}u_{-}(p),
\end{equation}
where $u_{-}(p)$ is a negative helicity spinor satisfying the massless
Dirac equation with momentum $p$. Using the freedom in choosing
the supersymmetry parameter $\eta$, we choose it to be a  negative
helicity spinor obeying the massless Dirac equation with arbitrary
momentum $k$ times a Grassmann variable $\theta$. Then,
\begin{equation}
\Gamma^+(p,k)\equiv\Gamma^+(p,\eta(k))=\theta\l k+|p-\r.
\end{equation}

As an application, we show how to derive the $b_{4;0}^{{\rm SUSY}}$
amplitudes from the well-known results for $d_{4;0}$ (see eq.\
(\ref{treegluon})). To obtain $\l \lambda_1^-\lambda_2^-\lambda_3^+
\lambda_4^+\r_0$, consider $\l [Q,g_1^-g_2^-g_3^+\lambda_4^+]\r_0$ and
$\l [Q,\lambda_1^-g_2^-\lambda_3^+\lambda_4^+]\r_0$:
\begin{eqnarray}
\lefteqn{0=\l [Q,g_1^-g_2^-g_3^+g_4^+]\r_0
=\Gamma^-(p_1,k)\l \lambda_1^-g_2^-g_3^+\lambda_4^+\r_0} \\ \nonumber
&&+\Gamma^-(p_2,k)\l g_1^-\lambda_2^-g_3^+\lambda_4^+\r_0
-\Gamma^-(p_4,k)\l g_1^-g_2^-g_3^+g_4^+\r_0,
\end{eqnarray}
\begin{eqnarray}
\lefteqn{0=\l [Q,\lambda_1^-g_2^-\lambda_3^+\lambda_4^+]\r_0
=-\Gamma^-(p_2,l)\l \lambda_1^-\lambda_2^-\lambda_3^+\lambda_4^+\r_0} \\
\nonumber
&&+\Gamma^-(p_3,l)\l \lambda_1^-g_2^-g_3^+\lambda_4^+\r_0
-\Gamma^-(p_4,l)\l \lambda_1^-g_2^-\lambda_3^+g_4^+\r_0.
\end{eqnarray}
Choose $k=p_2$ and $l=p_4$. Using the properties of the spinor
products and momentum conservation, we find
\begin{equation}
\l \lambda_1^-\lambda_2^-\lambda_3^+\lambda_4^+\r_0\equiv
b_{4;0}^{{\rm SUSY}}(-,-,+,+)=-\i\frac{\l 12\r[34]}{s_{14}}.
\end{equation}
In an exactly analogous fashion, from $\l [Q,g_1^-g_2^+g_3^-
\lambda_4^+]\r_0$ and $\l [Q,\lambda_1^-g_2^+\lambda_3^-\lambda_4^+]\r_0$
with $k=p_3$ and $l=p_4$, we obtain
\begin{equation}
\l \lambda_1^-\lambda_2^+\lambda_3^-\lambda_4^+\r_0\equiv
b_{4;0}^{{\rm SUSY}}(-,+,-,+)=
\i\frac{\l 13\r[24]}{s_{14}}\frac{s_{13}}{s_{12}}.
\end{equation}

We now see that our results given in eq.\ (\ref{SUSYresultDR}) are
indeed supersymmetric due to the universality of the $F_1^{-\pm}$
functions. When applying a supersymmetry transformation,
the only change one finds is such that the Born terms transform into
one another.

{}From the results of sect.\ 5, we see that the color subamplitudes
calculated in the \HV and \DR schemes only slightly differ.
Consequently, one expects that the \HV scheme breaks supersymmetry
only by a small amount at one-loop level. This is indeed what one
finds in a direct calculation:
\begin{equation}
\label{SUSYresultHV}
m^{{\rm SUSY,HV}}_{4;1}(-,\mp,\pm,+)=
c_\Gamma m^{{\rm SUSY}}_{4;0}(-,\mp,\pm,+)
N_c\left(F^{-\mp}_1(\varepsilon, s_{12}, s_{13}, s_{14})
-\frac{\delta_m}{3}\right).
\end{equation}
with the same process independent $F^{-\mp}_1$ functions as in eq.\
(\ref{SUSYresultDR}), but with process dependent $\delta_m$:
$\delta_b=5$, $\delta_c=3$, $\delta_d=1$.

According to the \SWI\ (\ref{SWI}), have we used a
supersymmetry-preserving regulator, the constants $\delta_m$ would have
been process independent in eq.\ (\ref{SUSYresultHV}).
In the followings, we shall argue that the supersymmetry breaking in
the \HV scheme can consistently be restored applying a special form
of finite renormalization.  We obtain this renormalization via
calculating the gluon and gluino self energies to
${\cal O}(\varepsilon)$ order in \HV scheme and \DR scheme and
require that
the finite difference between the results can be compensated by a
finite renormalization. For the self energies, we have the well-known
expressions
\begin{equation}
\Pi^{ab}_{\mu\nu}(k)=
\i \delta^{ab}(k_\mu k_\nu-k^2g_{\mu\nu})\Pi(k^2),
\end{equation}
\begin{equation}
\i \Sigma^i_j(k)=\delta^i_j\Sigma(k^2)\slash{k}.
\end{equation}
The functions $\Pi(k^2)$ and $\Sigma(k^2)$ are regularization-scheme
dependent.  Explicit calculation for space-like $k$ in the
background-field gauge gives
\begin{eqnarray}
\Pi^{{\rm HV}}(k^2)&=&N_c\left(\frac{g_r}{4\pi}\right)^2
\left(-\frac{4\pi\mu^2}{k^2}\right)^\varepsilon\Gamma(1+\varepsilon)
\left(-\frac{3}{\varepsilon}-\frac{57}{9}\right)+Z^f_A-1
+{\cal O}(\varepsilon),\\
\vspaceinarray
\Pi^{{\rm DR}}(k^2)&=&N_c\left(\frac{g_r}{4\pi}\right)^2
\left(-\frac{4\pi\mu^2}{k^2}\right)^\varepsilon\Gamma(1+\varepsilon)
\left(-\frac{3}{\varepsilon}-\frac{54}{9}\right)
+{\cal O}(\varepsilon),\\
\vspaceinarray
\Sigma^{{\rm HV}}(k^2)&=&N_c\left(\frac{g_r}{4\pi}\right)^2
\left(-\frac{4\pi\mu^2}{k^2}\right)^\varepsilon\Gamma(1+\varepsilon)
\left(\frac{1}{\varepsilon}+1\right)+Z^f_\lambda-1
+{\cal O}(\varepsilon),\\
\vspaceinarray
\Sigma^{{\rm DR}}(k^2)&=&N_c\left(\frac{g_r}{4\pi}\right)^2
\left(-\frac{4\pi\mu^2}{k^2}\right)^\varepsilon\Gamma(1+\varepsilon)
\left(\frac{1}{\varepsilon}+2\right)
+{\cal O}(\varepsilon).
\end{eqnarray}
Requiring that the results are equal in the two schemes, we obtain the
desired finite renormalizations:
\begin{equation}
Z^f_A=1+\frac{1}{3}N_c\left(\frac{g_r}{4\pi}\right)^2
\end{equation}
and
\begin{equation}
Z^f_\lambda=1+N_c\left(\frac{g_r}{4\pi}\right)^2
\end{equation}
So far, this renormalization is very similar to that of sect.\ 6,
where we discussed the difference between \lQCD\ in the \HV and \DR
schemes. There is, however, an important difference. The
renormalizations discussed in the present section have to
vanish as $k^2\ra 0$.  The reason is the following:
In the spirit of dimensional regularization, the \UV and \IR
divergencies cancel exactly for on-shell propagation independently
whether \HV scheme or \DR has been used.
Therefore, the only breaking of supersymmetry could occur when
calculating the loop correction for off-shell lines. This implies that
the renormalization insertions occur on off-shell
propagator lines only. Of course, in order not to break gauge
invariance, additional finite renormalization of the vertices is
necessary according to the Ward identities
\begin{equation}
Z^f_{AAA}=Z^f_g(Z^f_A)^{3/2},\;
Z^f_{A\lambda\lambda}=Z^f_g(Z^f_A)^{1/2}Z^f_\lambda,
\end{equation}
where $Z^f_g$ is the gauge-coupling renormalization\footnote{We note
that in the helicity formalism the renormalization of the four-gluon
vertex at the one-loop level is irrelevant.}.
{}From these equations we obtain the following condition among the
renormalization constants
\begin{equation}
\label{Zcondition}
Z^f_{AAA}(Z^f_A)^{-1}= Z^f_{A\lambda\lambda}(Z^f_\lambda)^{-1}.
\end{equation}
This condition is sufficient to ensure that if the renormalization
procedure is consistent for the four-point amplitudes then it is
consistent for any $n$-point one-loop amplitude. The proof goes by
induction. We assume that the finite renormalization consistently
restores supersymmetry for the $(n-1)$-point amplitudes. Attaching an
external gluon leg on an $(n-1)$-point tree amplitude on a gluon line
(either propagator or external leg) we introduce a finite
renormalization of $Z^f_{AAA}(Z^f_A)^{-1}$, while attaching the
external gluon on a fermion line we introduce a renormalization of
$Z^f_{A\lambda\lambda}(Z^f_\lambda)^{-1}$. According to eq.\
(\ref{SUSYresultHV}), these renormalizations
are equal, therefore, if the renormalization was consistent at
$(n-1)$-point level, it remains consistent at $n$-point level. Using
this argument, and the results of ref.\ \cite{Ber93} for the
five-gluon amplitudes in the pure gauge sector presented in the
\HV scheme, we can deduce the field theory result in dimensional
reduction which turns out to be identical to the result in the \FDH
scheme. This is a further indication that \DR in field theory gives
the same result as the \FDH scheme when the string based rules are
used.

The consistency of our renormalization procedure at the four-point
level can be seen by performing the finite renormalization on
formula (\ref{SUSYresultHV}) explicitly. We find that the constants
$\delta_m$ in eq.\ (\ref{SUSYresultHV}) become universal, hence the
required \SWI\ are fulfilled. However, the absolute normalization is
not set by condition (\ref{Zcondition}), consequently, the actual
value of $\delta_m$ is unknown. In order to set the absolute
normalization, we require the background-gauge Ward identity between
the gauge-field and gauge-coupling renormalizations\footnote{We use the
background field formalism to the gluon sector only, therefore, we
have to use the non-supersymmetric condition.} \cite{Abb81}, namely
\begin{equation}
Z^f_g=(Z^f_A)^{-1/2}=
1-\frac{1}{6}N_c\left(\frac{g_r}{4\pi}\right)^2.
\end{equation}
Having the absolute normalization set, we obtain $\delta_m=0$, i.e.,
we recover the result obtained using \DR exactly together with the
correct shift in \lQCD\ between the two schemes.

\section{Bremsstrahlung contributions in \HV and \DR schemes}
\setcounter{equation}{0}

In the first part of this paper, we have seen that one-loop amplitudes
of massless QCD can be calculated using the \HV version of dimensional
regularization and a simple rule for the transition to the \CDR scheme
has been found. The advantage of using the \HV scheme is obvious: both
the calculation and the final result is much simpler than in the \CDR
scheme. We have also
seen the usefulness of performing the loop calculation using a
supersymmetry preserving or weakly violating regulator. On the other
hand, the most widely accepted regularization scheme for the
regularization of \IR divergencies which is also used in the
factorization procedure is the \CDR scheme. These facts motivate a
thorough study of the various dimensional regularization schemes to
establish the transition rules among them. We already know, how to make
the transition for the loop corrections. In order to establish the
transition for physical quantities, we have to investigate the
Bremsstrahlung contributions as well.

To start with, we recall the form of the Bremsstrahlung contribution
found in ref.\ \cite{Kun92}. In order to keep the discussion reasonably
concise, we give the integrals in a schematic form and spell out those
factors explicitly which depend on the regularization scheme.  We
admit that it is difficult to follow our discussion without consulting
ref.\ \cite{Kun92} for more details and precise definitions. However,
we find it pointless to recapitulate every details which can be found
in ref.\ \cite{Kun92}.

The structure of the real corrections after integrating over the
phase space is same as eq.\ (\ref{Kunsztform}) for the virtual
corrections, although the actual expression is far more complicated.
Schematically we write
\begin{equation}
\I[2\ra3]=\sum_{i=1}^4\I[2\ra3]_i,\;
\I[2\ra3]_i=\I[{\rm soft}]_i+\I[{\rm coll}]_i+\I[{\rm NS}]_i,
\end{equation}
where $\I[2\ra3]$ is the $[2\ra3]$ cross section. To obtain $\I[2\ra3]$,
we have to integrate over the momenta of the three final state partons.
We can always arrange the calculation such that we label by 3 the parton
with the smallest transverse momentum.  The squared matrix element can
be decomposed into a sum of four terms, where one term $\l|\A_5^{{\rm
tree}}|^2\r_i$ ($i=1,2,3,4$) has singularities only when parton 3 is
soft or collinear to parton $i$.  Then $\I[2\ra3]_i$ is the part of the
cross section obtained by integrating $\l|\A_5^{{\rm tree}}|^2\r_i$;
$\I[{\rm soft}]_i$, $\I[{\rm coll}]_i$ and $\I[{\rm NS}]_i$ are its
soft, collinear and finite terms respectively.  The actual form of the
$\I[{\rm soft}]_i$, $\I[{\rm coll}]_i$ and $\I[{\rm NS}]_i$ integrals
depends whether parton $i$ is in the initial or final state.

\subsection{Soft contributions}

In the soft limit $p_3\ra 0$, the variables of parton 3 can be
integrated analytically. One is left with an integral over the variables
of partons 1 and 2:
\begin{equation}
\label{softint}
\I[{\rm soft}]_i=
\sum_{a,b,j_1,j_2}\int D_i L(a,b) \S_2(p_{j_1}^\mu,p_{j_2}^\mu)
\psi^{{\rm soft}}_i(\vec{a},\vec{p}),
\end{equation}
where the sum runs over all possible flavors of the initial
($a$ and $b$) and final ($j_1$ and $j_2$) state partons,
$D_i$ is a ubiquitous prefactor which does not depend on the
regularization scheme. The function $L(a,b)$ describes the parton
luminosity, $\S_2(p_{j_1}^\mu,p_{j_2}^\mu)$ is the measurement
function and the function $\psi^{{\rm soft}}_i(\vec{a},\vec{p})$ is
the soft limit of the squared matrix element integrated over the
variables of parton 3. They all depend on the regularization scheme.
However, the dependence in the parton luminosity and the measurement
function is irrelevant as far as our considerations are
concerned because the same type of integral occurs when the $[2\ra2]$
matrix elements are integrated over the phase space (see eqs.\ (29)
and (33) in ref.\ \cite{Kun92}).

When $i$ is a final state parton, the function
$\psi^{{\rm soft}}_i(\vec{a},\vec{p})$ has the form
\begin{eqnarray}
\label{finsoft}
\psi^{{\rm soft}}_i(\vec{a},\vec{p})&=&
\psi^{(4)}(\vec{a},\vec{p})\left\{\frac{1}{\varepsilon^2}C(a_i)-
\frac{1}{\varepsilon}2C(a_i)\log(x)\right\} \\ \nonumber
&&+\sum_{m\ne i}\psi^{(4,c)}_{im}(\vec{a},\vec{p})
\left\{-\frac{1}{2\varepsilon}\log\left(\frac{2p_i\cdot p_m}{Q^2}\right)
+\tilde{\I}_{im}(x)+{\cal O}(\varepsilon)\right\},\qquad (i=1,2),
\end{eqnarray}
while for the case when $i$ is an inital state parton, we have
\begin{eqnarray}
\label{inisoft}
\psi^{{\rm soft}}_i(\vec{a},\vec{p})&=&
\psi^{(4)}(\vec{a},\vec{p})\left\{\frac{1}{\varepsilon^2}C(a_i)+
\frac{1}{\varepsilon}2C(a_i)
\log\left(\frac{x_i}{1-x_i}\right)\right\} \\ \nonumber
&&+\sum_{m\ne i}\psi^{(4,c)}_{im}(\vec{a},\vec{p})
\left\{-\frac{1}{2\varepsilon}\log\left(\frac{2p_i\cdot p_m}{Q^2}\right)
+\tilde{\I}_{im}(x)+{\cal O}(\varepsilon)\right\},\qquad (i=3,4).
\end{eqnarray}
There is a term proportional to $-(1/\varepsilon)\log(x)$ in eq.\
(\ref{finsoft}) (the same $x$ appears in the argument of the
$\tilde{\I}_{im}$ functions too). In this term, $x$ is an arbitrary
positive number, less than one and it represents the fraction of $p_i$
that sets the upper limit of the integration in $p_3$. It is used to
separate the soft and collinear contributions in the soft-collinear
region. The limit $x\ra1$ means that in the collinear region all
momentum configuration is considered collinear except when
$p_3\simeq 0$. In ref.\ \cite{Kun92}, $x=1/2$ has
been chosen for the sake of convenience. For our purpose of establishing
the transition rules among the regularization schemes, it is more
suitable to keep $x$ arbitrary. The reason will be clear when the
collinear integrals are investigated.

Using the derivations in Appendix A of ref.\ \cite{Kun92}, we find
that in eqs.\ (\ref{finsoft}) and (\ref{inisoft}) only the $\psi^{(4)}$
and $\psi^{(4,c)}_{im}$ functions depend on the regularization scheme:
for the \CDR scheme they are the $d$-dimensional expressions, while for
the \HV or \DR schemes they are to be taken in four dimensions.

\subsection{Collinear contributions}

The integral for the final state collinear singularities has an
analogous form to that of the soft terms in eq.\ (\ref{softint}):
\begin{equation}
\label{fincoll}
\I[{\rm coll}]_i=
\sum_{a,b,j_1,j_2}\int D_i L(a,b) \S_2(p_{j_1}^\mu,p_{j_2}^\mu)
\psi^{{\rm coll}}_i(\vec{a},\vec{p}),\qquad (i=1,2).
\end{equation}
In eq.\ (\ref{fincoll}),
\begin{equation}
\psi^{{\rm coll}}_i(\vec{a},\vec{p})=-\frac{1}{\varepsilon}
\left(\frac{Q^2}{16p_i^2}\right)^\varepsilon
\frac{\Gamma(1-\varepsilon)}{\Gamma(1+\varepsilon)\Gamma(1-2\varepsilon)}
{\cal Z}(a_i,x) \psi^{(4)}(\vec{a},\vec{p})
\end{equation}
is the collinear limit of the squared matrix element integrated over
the variables of parton 3. The function ${\cal Z}(a,x)$ is an
integral of a finite expression:
\begin{equation}
{\cal Z}(a,x)=\int_x^1 \d z(1-z)^{-2\varepsilon}\left[z^{-2\varepsilon}
\sum_b \tilde{P}_{b/a}(z,\varepsilon)-\frac{2C(a)}{1-z}\right].
\end{equation}
The lower limit of the integration is the same $x$ we have met in eq.\
(\ref{finsoft}). $\tilde{P}_{b/a}(z,\varepsilon)$ is the usual
\AP splitting function in ($4-2\varepsilon$) dimensions without $z\ra1$
regulation. We shall term them $\varepsilon$-dependent \AP kernels.
In Appendix B, we collected these $\varepsilon$-dependent \AP kernels
in all three regularization schemes. Using
those expressions, one can easily verify that ${\cal Z}(a,x)$ does not
depend on the scheme in the limit $x\ra 1$. We are allowed to take that
limit, because the sum of the soft and collinear terms is independent of
$x$ (as can be checked using the precise formul\ae\ of ref.\
\cite{Kun92}).  Therefore, we find again that in the final state
collinear term, eq.\ (\ref{fincoll}), apart from the irrelevant
dependence in $L$ and $\S_2$, it is only the Born term,
$\psi^{(4)}(\vec{a},\vec{p})$ which is regularization-scheme dependent.

The structure of $\I[{\rm coll}]_i$ for initial state parton is
more complicated. It can be split into two terms. The first has the
structure encountered at the final state terms, while the second
contains the the initial state collinear pole that is involved in the
factorization theorem\footnote{In ref.\ \cite{Kun92}, the whole
contribution has been called $G^{{\rm coll}}_i$. For our purposes this
decomposition is useful.}:
\begin{equation}
\label{inicoll}
\I[{\rm coll}]_i=
\sum_{a,b,j_1,j_2}\int \left[D_i L(a,b) \S_2(p_{j_1}^\mu,p_{j_2}^\mu)
\psi^{{\rm coll}}_i(\vec{a},\vec{p})
+G^{{\rm coll}}_i(\vec{a},\vec{p})\right],\qquad (i=3,4).
\end{equation}
In eq.\ (\ref{inicoll}), $\psi^{{\rm coll}}_i$ has the form
\begin{eqnarray}
\psi^{{\rm coll}}_i(\vec{a},\vec{p})&=&
\psi^{(4)}(\vec{a},\vec{p})\left\{\frac{1}{\varepsilon}
\left[\gamma(a_i)-2C(a_i)\log\left(\frac{x_i}{1-x_i}\right)\right]
\right. \\ \nonumber
&&+\left.\log\left(\frac{Q^2}{\mu^2}\right)
\left[\gamma(a_i)-2C(a_i)\log\left(\frac{x_i}{1-x_i}\right)\right]
\right\}.
\end{eqnarray}
Again, the whole regularization-scheme dependence in $\psi^{{\rm
coll}}_i$ is contained in the Born term. The other term, $G^{{\rm
coll}}_i$ has a more complicated structure. As always, we emphasize only
that part which depends on the regularization scheme,
\begin{eqnarray}
\label{Gcoll}
G^{{\rm coll}}_i(\vec{a},\vec{p})&=&-G_{{\rm CT},i}^{(2\ra2)} \\
\nonumber &&-\alpha_s^2\frac{p_2}{2s^2}
\left(\frac{p_2}{2\pi\mu}\right)^{-2\varepsilon}\!\!
\psi^{(4)}(\vec{a},\vec{p})\S_2(p_{j_1}^\mu,p_{j_2}^\mu)
\sum_{\tilde{a}}\frac{\omega(\tilde{a})}{\omega(a_i)}
\int_{x_i}^1\frac{\d z}{z^2}L\left(\tilde{a},b,\frac{x_i}{z}\right)
\frac{\alpha_s}{2\pi}
\tilde{P}'_{a_i/\tilde{a}}(z) \\ \nonumber
&&+{\rm finite\: terms}.
\end{eqnarray}
In eq.\ (\ref{Gcoll}), $G_{{\rm CT},i}^{(2\ra2)}$ is the part of the
\ms\ factorization counter term that is associated with particle $i$,
\begin{eqnarray}
\label{mscounterterm}
\lefteqn{G_{{\rm CT},i}^{(2\ra2)}=} \\ \nonumber
&&\alpha_s^2\frac{p_2}{2s^2}
\left(\frac{p_2}{2\pi\mu}\right)^{-2\varepsilon}\!\!
\psi^{(4)}(\vec{a},\vec{p})\S_2(p_{j_1}^\mu,p_{j_2}^\mu)
\sum_{\tilde{a}}\frac{\omega(\tilde{a})}{\omega(a_i)}
\int_{x_i}^1\frac{\d z}{z^2} L\left(\tilde{a},b,\frac{x_i}{z}\right)
\frac{(4\pi)^\varepsilon}{\varepsilon\Gamma(1-\varepsilon)}
\frac{\alpha_s}{2\pi}\tilde{P}'_{a_i/\tilde{a}}(z)
\end{eqnarray}
It cancels exactly, when the factorization counter term is added.
The Born term and the factors $\omega(a)$
which represent the number of spin and color states of a parton type
$a$ are four-dimensional expressions in the second line of eq.\
(\ref{Gcoll}). $\tilde{P}'$ represents the $\varepsilon$-dependent
part of the $\varepsilon$-dependent \AP kernels,
\begin{equation}
\tilde{P}'_{a/b}(z)=\frac{\partial}{\partial\varepsilon}
\tilde{P}_{a/b}(z,\varepsilon)\lower 4pt\hbox{$|_{\varepsilon=0}$}.
\end{equation}
The functions $\tilde{P}'$ are given explicitly in Appendix B.
The scheme dependence in eq.\ (\ref{Gcoll}) occurs in the $\tilde{P}'$
functions only.

\bigskip

Summarizing this section, we conclude that the dependence on the
regularization scheme in $\I[2\ra3]$ appears only in the $\psi^{(4)}$
and $\psi^{(4,c)}_{im}$ functions and in a folding of the parton
luminosities with the $\varepsilon$-dependent part of the
$\varepsilon$-dependent \AP kernels, $\tilde{P}'$. The transition
rule from the \CDR scheme to the \HV or \DR schemes is very simple:
the $\psi^{(4)}$ and $\psi^{(4,c)}_{im}$ functions has to be taken in
four dimensions and $L\otimes\tilde{P}'$ has to be chosen in the proper
scheme according to the formulas in Appendix B. This transition rule
is in accordance with the transition rule found in sect.\ 6 for the
virtual contributions.

\section{Transition rules for the hard-scattering cross section among
dimensional regularization schemes}
\setcounter{equation}{0}

According to the factorization theorem, the physical cross section in
the QCD improved parton model for hadron-hadron scattering is a
folding between the parton densities and the hard-scattering cross
section:
\begin{equation}
\d\sigma(p_A,p_B)=\sum_{a,b}\int_0^1\d x_a\int_0^1\d x_b
f_{a/A}(x_a,\mu)f_{b/B}(x_b,\mu)\d\hat{\sigma}_{a,b}(x_a p_A, x_b p_B,
\mu, \alpha_s(\mu)).
\end{equation}
In this equation, the $f_{a/A}$ parton densities are process
independent, their $Q^2$-evolution is determined by the \AP equations.
At \NLO\ the regularization and factorization scheme dependence of
the \AP kernels is cancelled by that of the hard-scattering cross
section, $\d\hat{\sigma}_{a,b}$. Therefore, in principle all schemes
are equally acceptable.  In practice, however, one is forced to use
the existing parton density functions which are obtained by fitting
a large set of data in a rather complicated phenomenological procedure.
This fitting and the $Q^2$-dependence of the fitted function is always
worked out using the \CDR regularization and \ms\ (or DIS)
factorization schemes. As a result, the existing parton density
functions can be used only with those hard-scattering cross sections
which are calculated in the \CDR dimensional regularization scheme.
On the other hand, as we have seen above, convenient applications of
the helicity method to loop calculations require that regularization
is performed in the \HV or \DR schemes. Therefore, it is vital to find
the transition rules which tell us how to transform the expressions for
the hard-scattering cross sections obtained in the \HV or \DR schemes
into the corresponding expressions in the \CDR scheme. That goal is
easily achievable using the results of sections\ 6 and 8.

The cancellation of \IR divergencies can be seen manifestly when the
analytic structure of the \IR singularities are exhibited according
to eqs.\ (\ref{Kunsztform}) and (\ref{softint}--\ref{mscounterterm}).
In this cancellation the $\psi^{(4)}(\vec{a},\vec{p})$ and
$\psi^{(4,c)}_{mn}(\vec{a},\vec{p})$ functions appear as formal
objects, the only requirement is that they are the same in the
real and virtual contributions and in the factorization counter term.
We have seen that it is indeed the case in all three versions of
dimensional regularization: in the \CDR scheme, they are $d$-dimensional
expressions, while in the \HV and \DR schemes, they are taken in four
dimensions.

After cancellation of \IR divergencies and subtraction of
initial-state collinear singularities, we find still process independent
differences in the hard-scattering cross sections in the different
regularization schemes. If we compare $\d\hat{\sigma}_{a,b}$ in the
\HV scheme to that in the \CDR scheme, then we find that only the
$\tilde{P}'$ functions are different. If we compare
$\d\hat{\sigma}_{a,b}$ in the \DR scheme to that in the \CDR scheme,
then we find that the difference is contained in part in the
redefinition of \lQCD, in part in the $\tilde{P}'$ functions
and in part in $\gamma(a,\varepsilon)$: in the
\CDR scheme $\gamma$ takes its $\varepsilon$-dependent value, while
in dimensional reduction, its $\varepsilon$-dependent part is set to
zero. From these considerations we see that the only modification
one has to perform on $\d\hat{\sigma}_{a,b}$ obtained in non-\CDR
schemes in order to recover $\d\hat{\sigma}_{a,b}$ in the \CDR scheme
is simply changing $\tilde{P}'$ to the value in the \CDR scheme,
use $\varepsilon$-dependent $\gamma$ parameters and the standard
$\Lambda_{\overline{{\rm MS}}}$ QCD parameter.

In order to find the proper form of the loop contributions in the \HV or
\DR schemes which is useful for the cancellation mechanism (i.e.,
eq.\ (\ref{Kunsztform})), we have to calculate the
$\psi^{(4)}(\vec{a},\vec{p})$ and $\psi^{(4,c)}_{mn}(\vec{a},\vec{p})$
functions in four dimensions. But these are tree level calculations
which can easily be done using the helicity method. For the
definition of the $\psi^{(4,c)}_{mn}(\vec{a},\vec{p})$ functions,
we refer to ref.\ \cite{Kun92}. Knowing the
$\psi^{(4)}(\vec{a},\vec{p})$ and $\psi^{(4,c)}_{mn}(\vec{a},\vec{p})$
functions in four dimensions and the complete loop contributions
in $d$ dimensions, we can
deduce the form of the $\psi^{(6)}_{{\rm NS}}(\vec{a},\vec{p})$
functions which constitute the finite part of the loop corrections in
the \CDR scheme. As far as the Bremsstrahlung contributions are
concerned, in the \HV and \DR schemes, both the soft and the hard part
of the cross section have to be evaluated in four dimensions. It is only
the calculation of the $\tilde{P}'$ functions in the \HV scheme which
requires $d$-dimensional treatment (see Appendix B).

We have given the transition terms and functions needed to recover the
hard-scattering cross section in the \CDR regularization scheme from
calculations performed in the \HV or \DR schemes. This is the main
result of our paper, which was based upon the
assumption that the analytic structure of the singularities found in
ref.\ \cite{Kun92} is process independent even after the minor
modification obtained by the introduction of $\varepsilon$-dependent
$\gamma$ parameters. One expects that this structure is indeed
universal because the singularities are linked to the external
particles of the subprocess. To give more support
to the universality, in sect.\ 10, we present further consistency
check on the transition rules via going into the supersymmetric limit
of QCD once again.

\section{Next-to-leading order \AP kernels in the \HV and \DR schemes}
\setcounter{equation}{0}

We have seen that the only dependence on the regularization scheme in
the hard-scattering cross section, $\d\hat{\sigma}_{a,b}$ --- beyond
the redefinition of \lQCD\ --- is contained in $\tilde{P}'$ and
$\tilde{\gamma}(a)$. We have established the transition rules to
recover the hard-scattering cross section in the \CDR \ms\ scheme from
a calculation performed in non-\CDR schemes.
Although, all phenomenological studies for the parton densities
are performed using the \CDR \NLO\ \AP kernels, it may still be useful
to know these kernels also for \DR and \HV schemes. In particular,
knowing the \NLO\ \AP kernels for a supersymmetry preserving
regularization, such as dimensional reduction, has got its own
interest.

The \NLO\ \AP kernels in  \DR can be obtained in two different ways.
On one hand, they can be determined in a direct calculation using
dimensional reduction. On the other, requiring that the physical
cross section should not depend on the regularization scheme, the
regularization-scheme dependence of the \NLO\ \AP kernels can be
uniquely determined from that of the hard-scattering cross section.
If the two approaches lead to the same result, then it
would prove that unitarity is not violated as well as it
would ensure the universality of our transition rules. A direct
calculation of the \NLO\ \AP kernels in non-\CDR schemes has not
been carried out yet. Nevertheless, we can still perform for the
\DR scheme a consistency test: we can test explicitly whether the
\NLO\ \AP kernels satisfy the \SWI\ for \N1 \YM theory.

For the sake of completeness, we review the derivation of the
transformation rule for the \NLO\ \AP kernels using the example
of deep-inelastic lepton-hadron scattering. The $F_2$ structure
function at \NLO\ is
\begin{equation}
\label{DISF2}
\frac{1}{x}F_2(x,Q)=\sum_i e_i^2\left(f_{i/A}(x,\mu)+
\frac{\alpha_s}{2\pi}
\sum_a f_{a/A}(z,\mu)\otimes_x C_{ia}(z,Q/\mu)\right)
+{\cal O}(\alpha_s^2),
\end{equation}
where the following notation has been introduced:
\begin{equation}
f\otimes_x g = \int_0^1\d y\,\d z\,f(z)g(y)\delta(x-yz)
\end{equation}
and $i=q,\bar{q}$, $a=q,g$.
In eq.\ (\ref{DISF2}), $C_{ia}$ denotes the hard-scattering cross
section --- their actual form is irrelevant for our purposes\footnote{
In the language of deep-inelastic scattering, $C_{ia}$ is usually
called coefficient function.}. We can
freely change $C_{ia}$ to $C_{ia}'=C_{ia}+\Delta C_{ia}$\footnote{The
DIS factorization scheme at \NLO\ is defined by $C_{ia}'=0$.} if we
simultaneously change the parton densities to
\begin{equation}
\label{newf}
f_{a/A}'(x,\mu)=f_{a/A}(x,\mu)-
\frac{\alpha_s}{2\pi} \sum_b f_{b/A}(\mu)\otimes_x
\Delta C_{ab}(Q/\mu)\raise-.5em\hbox{$|_{Q=\mu}$}.
\end{equation}
Indeed, writing eq.\ (\ref{DISF2}) in terms of $C_{ia}'$ and
$f_{a/A}'$, we find that the
change in $F_2$ is of the order ${\cal O}(\alpha_s^2)$ which is
neglected. The $\mu$ dependence of both $f_{a/A}$ and $f_{a/A}'$ is
determined by the \AP evolution equations, but with different kernels
$P$ and $P'$ respectively:
\begin{equation}
\mu\frac{\partial}{\partial\mu}f_{a/A}^{(')}(x,\mu)=
\frac{\alpha_s}{\pi}\sum_bP_{a/b}^{(')}\otimes_x f_{b/A}^{(')}(\mu)
\end{equation}
In the equation for $f_{a/A}'$, we substitute $f_{a/A}'$ according
to eq.\ (\ref{newf}) on the left hand side. Performing the
differentiation and using
\begin{equation}
\mu\frac{\partial\alpha_s}{\partial\mu}=
-\frac{1}{2}\beta_0\frac{\alpha_s}{\pi}\alpha_s,
\end{equation}
we obtain
\begin{eqnarray}
\lefteqn{\sum_b P'_{a/b}\otimes_x f'_{b/A}(\mu)=} \\ \nonumber
&&\sum_b \left[P_{a/b}\otimes_x f_{b/A}(\mu)+\frac{\alpha_s}{2\pi}
\left(\frac{\beta_0}{2}f_{a/A}\otimes_x \Delta C_{ab}
-\sum_c(P_{b/c}\otimes f_{c/A})\otimes_x \Delta C_{ab}\right)\right]
+{\cal O}(\alpha_s^2).
\end{eqnarray}
We can rewrite the right hand side in terms of $f'$ and read off
the change in the \NLO\ \AP kernel induced by a change in the
hard-scattering cross section:
\begin{equation}
\label{newP}
P'_{a/b}(z)=P_{a/b}(z)+\frac{\alpha_s}{2\pi}
\left[\sum_c
\left(P_{a/c}\otimes_z\Delta C_{cb}
-\Delta C_{ac}\otimes_zP_{c/b}\right)
+\frac{\beta_0}{2}\Delta C_{ab}\right]
+{\cal O}(\alpha_s^2).
\end{equation}

Table 1 contains the change in the hard-scattering cross section  for
making the transition from the \CDR scheme to the \HV and \DR
regularization schemes in the \ms\ factorization scheme. These values
can be obtained from the $\tilde{P}'$ and $\tilde{\gamma}(a)$ functions
as described in Appendix B.
Substituting these values for $\Delta C_{ab}$ into eq.\ (\ref{newP}),
we obtain the \NLO\ \AP kernels in the \ms\ scheme. We list below
only the difference in comparison to the standard \ms\ expressions
obtained in the \CDR scheme \cite{Fur80}:
\begin{eqnarray}
P^{\overline{{\rm HV}}}_{g/g}(z)-
P^{\overline{{\rm MS}}}_{g/g}(z)&=&
C_F T_R N_f (2-\frac{2}{3}x^{-1}-\frac{4}{3}x^2+2x\log{x})
+C_G \beta_0 x(1-x).\\
\vspaceinarray
P^{\overline{{\rm HV}}}_{q/g}(z)-
P^{\overline{{\rm MS}}}_{q/g}(z)&=&
C_F T_R N_f (-1+x+2x\log{x}-4x^2\log{x}) \\ \nonumber
&+&C_G T_R N_f
(-1+\frac{2}{3}x^{-1}-4x+\frac{13}{3}x^2-4x\log{x}+4x^2\log{x}).\\
\vspaceinarray
P^{\overline{{\rm HV}}}_{g/q}(z)-
P^{\overline{{\rm MS}}}_{g/q}(z)&=&
C_F C_G (2-\frac{2}{3}x^{-1}-\frac{4}{3}x^2+2x\log{x}).\\
\vspaceinarray
P^{\overline{{\rm HV}}}_{q/q}(z)-
P^{\overline{{\rm MS}}}_{q/q}(z)&=&
C_F T_R N_f (-2+\frac{2}{3}x^{-1}+\frac{4}{3}x^2-2x\log{x}).\\
\vspaceinarray
\vspaceinarray
\label{DRscheme1}
P^{\overline{{\rm DR}}}_{g/g}(z)-
P^{\overline{{\rm MS}}}_{g/g}(z)&=&
C_F T_R N_f (3-\frac{2}{3}x^{-1}+x-\frac{10}{3}x^2+4x\log{x}).\\
\vspaceinarray
P^{\overline{{\rm DR}}}_{q/g}(z)-
P^{\overline{{\rm MS}}}_{q/g}(z)&=&
C_F T_R N_f (-3+2x+x^2-\log{x}-4x^2\log{x}) \\ \nonumber
&+&C_G T_R N_f (-2+\frac{2}{3}x^{-1}-10x
+\frac{34}{3}x^2-8x\log{x})  \\ \nonumber
&+&T_R N_f (\tilde{\gamma}(g)-\tilde{\gamma}(q))(1-2x+2x^2).\\
\vspaceinarray
P^{\overline{{\rm DR}}}_{g/q}(z)-
P^{\overline{{\rm MS}}}_{g/q}(z)&=&
C_F^2 (-x^{-1}+\frac{5}{2}x-2\log{x}-2x\log{x}) \\ \nonumber
&+&C_F C_G (2-x^{-1}+x-2x^2+4x\log{x})\\ \nonumber
&+&C_F [-\beta_0 x
+(\tilde{\gamma}(g)-\tilde{\gamma}(q))(2-2x^{-1}-x)].\\
\vspaceinarray
\label{DRscheme4}
P^{\overline{{\rm DR}}}_{q/q}(z)-
P^{\overline{{\rm MS}}}_{q/q}(z)&=&
C_F T_R N_f (-\frac{9}{3}+\frac{2}{3}x^{-1}-x
+\frac{10}{3}x^2-4x\log{x}) \\ \nonumber
&+&C_F \beta_0 \frac{1}{2}(-1+x).
\end{eqnarray}
In these equations, we used the notation of ref.\ \cite{Fur80}:
$C_F=V/(2N_c)$, $C_G=N_c$ and $T_R=1/2$.

Now we turn to the \N1 \YM theory discussed in sect.\ 7.
In this case, if the regularization scheme is supersymmetric,
the \AP kernels are to satisfy the following identity to all
orders in perturbation theory:
\begin{equation}
\label{SUSYidentity}
P_{g/g}(z)+P_{\tilde{g}/g}(z)=
P_{g/\tilde{g}}(z)+P_{\tilde{g}/\tilde{g}}(z).
\end{equation}
The validity of this relation follows simply from the physical meaning
of the \AP kernels: they give the number density of the partons in the
infinite momentum frame.
Therefore, if we sum over the final states, it should not
matter whether the initial state is a gluon or a gluino. The \AP
kernels are known to satisfy identity (\ref{SUSYidentity}) to leading
\cite{Fur80} and in \DR to \NLO\ \cite{Ant81}.

Some care is to be taken when going to the supersymmetric limit of
eqs.\ (\ref{DRscheme1}--\ref{DRscheme4}). A virtual gluino loop on a
gluon propagator contains a combinatorical factor of 1/2, while the
gluon splitting into a gluino pair does not contain the same factor.
Therefore, to obtain the supersymmetric limit, we take $T_R N_f\ra
N_c$ in the splitting functions and $2T_R N_f\ra N_c$ in loop
corrections. This rule can be written more consistently by using
$T_R N_f\ra N_c$, $\beta_0\ra 3N_c$, $\tilde{\gamma}(g)\ra N_c/6$
and $\tilde{\gamma}(q)\ra N_c/2$ in addition to the usual
rules, $C_F\ra N_c$, $C_G\ra N_c$ (this is the reason for not writing
$\beta_0$, $\tilde{\gamma}(g)$ and $\tilde{\gamma}(q)$ in terms of
$C_G$, $T_R$ and $N_f$ in eqs.\ (\ref{DRscheme1}--\ref{DRscheme4})).
{}From eqs.\ (\ref{DRscheme1}--\ref{DRscheme4}) for the combination
\begin{equation}
\Delta(z)\equiv \hat{P}_{g/g}(z)+\hat{P}_{\tilde{g}/g}(z)-
\hat{P}_{g/\tilde{g}}(z)-\hat{P}_{\tilde{g}/\tilde{g}}(z),
\end{equation}
we obtain
\begin{equation}
\label{Deltadiff}
\Delta^{\overline{{\rm DR}}}(z)-\Delta^{\overline{{\rm MS}}}(z)=
-\left(\frac{5}{6}-\frac{23}{3}x+7x^2-(2x+4x^2-1)\log{x}
+\frac{2}{3}x^{-1}\right)N_c^2,
\end{equation}
where $\hat{P}_{a/b}(z)$ is the \AP kernel without the part that is
proportional to $\delta(1-z)$. Comparing eq.\ (\ref{Deltadiff}) to
formula (19) of ref.\ \cite{Fur80}, we conclude that in the \DR
scheme $\Delta(z)=0$. This is the consistency check on our transition
rules we referred to above. This test, however, still does not
prove that unitarity is maintained in the \HV and \DR schemes.
It may be violated by terms which are proportional to $(N_c - N_f)$.

Finally, we remark that our derivation of eq.\ (\ref{Deltadiff}) is
different from that of Antoniadis and Floratos \cite{Ant81} and can be
considered as an independent confirmation of the result that in
dimensional reduction, relation (\ref{SUSYidentity}) remains valid in
\NLO.

\section{Summary}

We calculated the one loop corrections to the helicity amplitudes
of all $2\ra 2$ parton scattering processes in QCD using dimensional
reduction as well as \HV regularization schemes for
regularizing both the ultraviolet and infrared singularities.
We explicitly demonstrated the cancellation of the soft and collinear
singularities between the loop corrections and Bremsstrahlung
contributions using the general structure of the Bremsstrahlung
singularities described in ref.\ \cite{Kun92}. Using somewhat
heuristic arguments concerning the differences between the results
obtained in the \HV scheme and in the \CDR dimensional regularization
scheme, we could reproduce the results of Ellis and Sexton for the
one loop contributions to the spin and color summed cross-sections
of all $2\ra 2$ parton subprocesses. Using the general form of the
soft and collinear singularities, we have found universal (process
independent) transition rules which can be used to transform the
results obtained in the \HV scheme and in \DR into those of the \CDR
regularization scheme.
These transformation rules have practical importance because
the phenomenological parton density functions determined from fits
to the data assume \NLO\ $Q^2$-evolution as given in the \CDR
dimensional regularization scheme. Therefore, when a physical
cross section is calculated, the hard scattering parton cross
sections have to be given as obtained in the conventional scheme.
Since we know how to translate the results between the above
mentioned three schemes, one can do the actual calculation in the
most convenient scheme. We note that in the \HV scheme and in \DR
one can use the helicity method directly.

With explicit calculation we demonstrated that in dimensional
reduction, as expected, the \SWI\ of the
scattering amplitudes of the $2\ra 2$ parton subprocesses of the
\N1 \YM theory are satisfied also in \NLO. QCD differs
from the \N1 \YM theory only in the color representation of the
quarks and gluinos, therefore, the \SWI\ can be used to obtain
a non-trivial test of the correctness of the calculation and to
obtain the amplitudes of different subprocesses with the help
of the \SWI.

We have shown that the supersymmetry Ward identities can be derived
also using the non-supersymmetric \HV regularization scheme
by working out simple finite renormalization factors.
Using this result we find that the string theory based \FDH scheme of
Bern and Kosower gives the same result for the four- and
five-gluon amplitudes as the dimensional reduction scheme.

It is not obvious that in the \HV scheme and in the dimensional
reduction unitarity is not violated by some finite terms when the soft
and collinear singularities are cancelled between the loop corrections
and the Bremstrahlung contributions. This difficulty  does not arise
if we know the transition rules to the conventional regularization
scheme. Using the transition rules we calculated the \NLO\ \AP kernels
for \DR and for the \HV scheme. Unitarity is maintained in
this case by construction.  We have found that the \NLO\ \AP
kernels of the \N1 \YM theory obtained in \DR satisfies
the \SWI\ \cite{Ant81}.

\bigskip

{\large \bf Acknowledgement}
{}~~~~We are greatful to Z. Bern, L. Dixon and D. Kosower for helpful
correspondence and for sending their paper on the loop integrals
prior to publication.

\renewcommand{\theequation}{A.\arabic{equation}}
\setcounter{equation}{0}
\newpage

\noindent\Large\bf Appendix A\normalsize

\bigskip\bigskip
In this Appendix, we spell out eq.\ (\ref{parametrized}) explicitly
for $N=2,3,4$ in a form which is used in the actual calculation.
We also calculate all the necessary integrals.

Case $N=2$:
\begin{equation}
\M=\frac{\i}{(4\pi)^2}\left(\frac{4\pi\mu^2}{-p^2}\right)^\varepsilon
\sum_{k=0}^1(-p^2)^k\Gamma(-k+\varepsilon)
\int_{[0,1]^2}\d x_1\,\d x_2\,\delta(1-x_1-x_2)(x_1 x_2)^{k-\varepsilon}
N_k(J,p),
\end{equation}
where
\begin{equation}
J_1=x_2 p,
\end{equation}
\begin{equation}
J_2=-x_1 p.
\end{equation}
$N_k(J,p)$ is a polynomial of $x_1$ and $x_2$. The general form of a
term in the integral is
\begin{equation}
\int_{[0,1]^2}\d x_1\,\d x_2\,\delta(1-x_1-x_2)x_1^{n_1} x_2^{n_2}
(x_1 x_2)^{k-\varepsilon}=B(n_1+k+1-\varepsilon,n_2+k+1-\varepsilon).
\end{equation}

Case $N=3$:
\begin{equation}
D(x,p)=-(x_1 x_2 p_2^2+x_2 x_3 p_3^2+x_3 x_1 p_1^2).
\end{equation}
For the $0\ra$ 4 partons case, two of the external legs are on shell.
If $i$ and $j$ are the indices of the propagators joining to the
massive external leg ($p_j$), then eq.\ (\ref{parametrized}) becomes
\begin{eqnarray}
\lefteqn{\M=
-\frac{\i}{(4\pi)^2}\left(\frac{4\pi\mu^2}{-p_j^2}\right)^\varepsilon
\sum_{k=0}^1(-p_j^2)^{k-1}\Gamma(1-k+\varepsilon)} \\ \nonumber
&&\int_{[0,1]^3}\d x_i\,\d x_j\,\d x_k\,\delta(1-x_i-x_j-x_k)
(x_i x_j)^{k-1-\varepsilon} N_k(J,p),
\end{eqnarray}
where
\begin{equation}
J_i=x_j p_j+x_k(p_j+p_k),
\end{equation}
\begin{equation}
J_j=x_k p_k+x_i(p_k+p_i),
\end{equation}
\begin{equation}
J_k=x_i p_i+x_j(p_i+p_j).
\end{equation}
$N_k(J,p)$ is a polynomial of $x_i$, $x_j$ and $x_k$. The general form
of a term in the integral is
\begin{equation}
\int_{[0,1]^3}\d x_i\,\d x_j\,\d x_k\,\delta(1-x_i-x_j-x_k)
x_i^{n_i} x_j^{n_j} x_k^{n_k} (x_i x_j)^{k-1-\varepsilon}.
\end{equation}
After changing variables,
\begin{eqnarray*}
x_i&=&x\\
x_j&=&y(1-x)\\
x_k&=&(1-x)(1-y),
\end{eqnarray*}
the integral becomes
\begin{eqnarray}
\lefteqn{\int_0^1\d x\int_0^1\d y\,x^{n_i}(1-x)^{n_j+n_k}
y^{n_j}(1-y)^{n_k}(x(1-x)y)^{k-1-\varepsilon}(1-x)=}\\
&&~~~~~~~~~~~~~~~~~~~
B(n_i+k-\varepsilon,n_j+n_k+k+1-\varepsilon)B(n_j+k-\varepsilon,n_k+1).
\end{eqnarray}

Case $N=4$:
\begin{equation}
D(x,p)=-(x_1 x_2 p_1^2+x_2 x_3 p_2^2+x_3 x_4 p_3^2+x_4 x_1 p_4^2+
x_1 x_3 s_{12}+x_2 x_4 s_{14}),
\end{equation}
where
\begin{equation}
s_{ij}=(p_i+p_j)^2.
\end{equation}
For the $0\ra$ 4-parton case, the external legs are on shell.
Therefore, eq.\ (\ref{parametrized}) becomes
\begin{eqnarray}
\lefteqn{\M=
\frac{\i}{(4\pi)^2}(4\pi)^\varepsilon
\sum_{k=0}^2\Gamma(2-k+\varepsilon)} \\ \nonumber
&&\mu^{2\varepsilon}\int_{[0,1]^4}\d x_1\,\d x_2\,\d x_3\,\d x_4\,
\delta(1-x_1-x_2-x_3-x_4)
\frac{N_k(J,p)}{(-x_1 x_3 s_{12}-x_2 x_4 s_{14})^{2-k+\varepsilon}}
\end{eqnarray}
where
\begin{equation}
J_1=x_2 p_1+x_3(p_1+p_2)+x_4(p_1+p_2+p_3),
\end{equation}
\begin{equation}
J_2=x_3 p_2+x_4(p_2+p_3)+x_1(p_2+p_3+p_4),
\end{equation}
\begin{equation}
J_3=x_4 p_3+x_1(p_3+p_4)+x_2(p_3+p_4+p_1),
\end{equation}
\begin{equation}
J_4=x_1 p_4+x_2(p_4+p_1)+x_3(p_4+p_1+p_2).
\end{equation}
$N_k(J,p)$ is a polynomial of $x_1$, $x_2$,
$x_3$ and $x_4$. The generic form of the integral is
\begin{eqnarray}
\lefteqn{\I^k[P({x_i})]=
(4\pi)^{-2+\varepsilon}\Gamma(2-k+\varepsilon)} \\ \nonumber
&&\times\mu^{2\varepsilon}\int_{[0,1]^4}
\d x_1\,\d x_2\,\d x_3\,\d x_4\,\delta(1-x_1-x_2-x_3-x_4)
\frac{P(\{x_i\})}{(-x_1 x_3 s_{12}-x_2 x_4 s_{14})^{2-k+\varepsilon}}.
\end{eqnarray}

It is remarkable that the same type of integrals, but only for $k=0$,
appear in the calculation of one-loop amplitudes using the
string-based approach \cite{Ber92}. In ref.\ \cite{BDK92}, a general
method has been described how to calculate a large class of these
one-loop Feynman parametric integrals. The main idea of the
evaluation is derivation of a differential relation between the basic
scalar integral (${\cal I}^k[1]$) and the so called scaled integrals
\cite{BDK92} which are related to ${\cal I}^k[P(x_i)]$ according to
\begin{equation}
\hat{\I}^k[\hat{P}({x_i})]\equiv\left(\prod_{j=1}^4\alpha_j\right)^{-1}
\I^k[P({x_i/\alpha_i})],
\end{equation}
where $\alpha_i$ are positive, real parameters, satisfying
$\alpha_1\alpha_3=(-s_{12})^{-1}$ and $\alpha_2\alpha_4=(-s_{14})^{-1}$.

The basic scalar integral can be evaluated either using the partial
differential equation technique of ref.\ \cite{BDK92}, or by direct
integration. We have chosen the latter approach and found agreement
with the result of ref.\ \cite{Ber92}. For the cases $k\ne 0$ the same
method with trivial modifications works.

\renewcommand{\theequation}{B.\arabic{equation}}
\setcounter{equation}{0}
\bigskip\bigskip

\noindent\Large\bf Appendix B\normalsize

\bigskip\bigskip

This Appendix is devoted to the derivation of the changes in the
coefficient functions appearing in Table 1.

The first step in the determination of the $\Delta C_{ab}$ functions
is the derivation of the $\varepsilon$-dependent \AP kernels in all
three regularization schemes. This accounts for finding the
$\varepsilon$-dependent splitting functions without $z\ra 1$
regulation. These functions in the \DR scheme coincide with the
four-dimensional \AP kernels first obtained by Altarelli and Parisi.
We recall them to set the notation:
\begin{eqnarray}
P_{g/g}(z)&=&2 C_G \left[\frac{z}{1-z}+\frac{1-z}{z}+z(1-z)\right], \\
P_{q/g}(z)&=&T_R N_f \left[z^2+(1-z)^2\right], \\
P_{g/q}(z)&=&C_F \frac{1+(1-z)^2}{z}, \\
P_{q/q}(z)&=&C_F \frac{1+z^2}{1-z}.
\end{eqnarray}
We denote the $\varepsilon$-dependent \AP kernels by the regularization
scheme-dependent functions $\tilde{P}_{a/b}(z,\varepsilon)$ and
decompose them into a scheme-independent and a scheme-dependent part:
\begin{equation}
\tilde{P}_{a/b}(z,\varepsilon)=
P_{a/b}(z)+\varepsilon\tilde{P}'_{a/b}(z).
\end{equation}
Our goal is to find the scheme-dependent functions,
$\tilde{P}'_{a/b}(z)$.

Clearly, when the initial parton is a quark, the only
dependence on the dimension of space-time enters in the integral over
the loop momentum. As a result, the splitting functions for quark
splitting are the same in the \HV and \CDR schemes. Repeating the
calculation of Altarelli and Parisi in ($4-2\varepsilon$) dimensions,
one finds \cite{Ell81}
\begin{eqnarray}
\label{CDRHVgq}
\tilde{P}'^{{\rm CDR}}_{g/q}(z)=
\tilde{P}'^{{\rm HV}}_{g/q}(z)&=&-C_F z,\\
\label{CDRHVqq}
\tilde{P}'^{{\rm CDR}}_{q/q}(z)=
\tilde{P}'^{{\rm HV}}_{q/q}(z)&=&-C_F (1-z).
\end{eqnarray}

We follow the procedure of Appendix A of ref.\ \cite{Kun92} to derive
the splitting functions for gluon splitting. In that reference, the
splitting for a polarized gluon were obtained in the form
\begin{eqnarray}
\label{polarizedAP}
\tilde{P}_{q/g}^{ij}(z, {\rm {\bf q}}, \varepsilon)&=&\frac{1}{2}
\left[\delta^{ij}-4z(1-z)\frac{q^iq^j}{{\rm {\bf q}}^2}\right], \\
\label{polarizedAP2}
\tilde{P}_{g/g}^{ij}(z, {\rm {\bf q}}, \varepsilon)&=&2 C_G
\left[\left(\frac{z}{1-z}+\frac{1-z}{z}\right)\delta^{ij}
+2(1-\varepsilon)z(1-z)\frac{q^iq^j}{{\rm {\bf q}}^2}\right].
\end{eqnarray}
In eqs.\ (\ref{polarizedAP}), (\ref{polarizedAP2}),
${\rm {\bf q}}$ is the transverse momentum of the
product particle, $q^i$ and $q^j$ are the two components of
${\rm {\bf q}}$. The $\varepsilon$-dependent \AP kernel is then
(see ref.\ \cite{Kun92})
\begin{equation}
\label{gluonAP}
\tilde{P}_{a/g}(z,\varepsilon)=
\frac{V}{\omega(g)}\delta_{ij}\tilde{P}_{a/g}^{ij}(z, {\rm {\bf q}}).
\end{equation}
Both $\omega(g)$ and the contraction of the transverse indices are
scheme dependent. In the \HV scheme they are taken in four dimensions,
while in \CDR dimensional regularization they are in $d$-dimensions.
Using eqs.\ (\ref{polarizedAP}-\ref{gluonAP}), we obtain
\begin{eqnarray}
\label{CDRgg}
\tilde{P}'^{{\rm CDR}}_{g/g}(z)&=&0,\\
\tilde{P}'^{{\rm CDR}}_{q/g}(z)&=&-2T_R N_f z(1-z)
\end{eqnarray}
in the \CDR  scheme, in agreement with the results of ref.\
\cite{Ell81}, while in the \HV scheme, we find
\begin{eqnarray}
\tilde{P}'^{{\rm HV}}_{g/g}(z)&=&-2C_G z(1-z),\\
\label{HVqg}
\tilde{P}'^{{\rm HV}}_{q/g}(z)&=&0,
\end{eqnarray}
in agreement with the results of ref.\ \cite{Gie92}.

The difference between the hard scattering cross sections when changing
the regularization scheme determines the
$\Delta C_{ab}(z)$ functions. These differences were analyzed in great
details in the main text. The differences can entirely be described by
the scheme dependence of the $\tilde{P}'$ functions and by that of the
$\tilde{\gamma}(a)$ constants. The $\gamma(a)$ constants represent
the contribution from the virtual graphs to the \AP kernel, therefore,
we may interpret the $\tilde{\gamma}(a)$ constants as the contribution
from the virtual graphs to the $\varepsilon$-dependent part of the
$\varepsilon$-dependent \AP kernel. The $\tilde{\gamma}(a)$ constants
are scheme dependent:
\begin{eqnarray}
\label{CDRgammatilde}
\tilde{\gamma}^{{\rm CDR}}(g)=\tilde{\gamma}^{{\rm HV}}(g)&=&
\frac{1}{6}N_c,\\
\tilde{\gamma}^{{\rm CDR}}(q)=\tilde{\gamma}^{{\rm HV}}(q)&=&
\frac{1}{2}C_F,
\end{eqnarray}
\begin{equation}
\label{DRgammatilde}
\tilde{\gamma}^{{\rm DR}}(g)=0,
\tilde{\gamma}^{{\rm DR}}(q)=0.
\end{equation}
Using eqs.\ (\ref{CDRHVgq}), (\ref{CDRHVqq}), (\ref{CDRgg}-\ref{HVqg}),
(\ref{CDRgammatilde}-\ref{DRgammatilde}), it is now easy to fill the
entries of Table 1.

\def\np#1#2#3  {{\it Nucl. Phys. }{\bf #1} (19#3) #2}
\def\nc#1#2#3  {{\it Nuovo. Cim. }{\bf #1} (19#3) #2}
\def\pl#1#2#3  {{\it Phys. Lett. }{\bf #1} (19#3) #2}
\def\pr#1#2#3  {{\it Phys. Rev. }{\bf #1} (19#3) #2}
\def\prep#1#2#3{{\it Phys. Rep. }{\bf #1} (19#3) #2}

\newpage


\newpage

\begin{thebibliography}{99}
\bibitem{Ber92} Z. Bern and D. A. Kosower, \np{B379}{451}{92} .
\bibitem{Ber93} Z. Bern L. Dixon, and D. A. Kosower, preprint
CERN-TH.6803/93.
\bibitem{Kun92} Z. Kunszt and D. E. Soper, \pr{D46}{192}{92} .
\bibitem{Cur80} G. Curci, W. Furmanski and R. Petronzio,
\np{B175}{27}{80} .
\bibitem{Gie92} W. T. Giele and E. W. N. Glover, \pr{D46}{1980}{92} .
\bibitem{Ber87} F. A. Berends and W. T. Giele, \np{B294}{700}{87} .
\bibitem{Lam69} C. S. Lam and J. P. Lebrun, \nc{59A}{397}{69} ;
C. S. Lam, preprint McGill/92-32.
\bibitem{Ell86} R. K. Ellis and J. C. Sexton, \np{B269}{445}{86} .
\bibitem{'tH72} G. 't Hooft and M. Veltman, \np{B44}{189}{72} .
\bibitem{Gas73} R. Gastmans and R. Meuldermans, \np{B63}{277}{73} .
\bibitem{Sie79} W. Siegel, \pl{84B}{193}{79} .
\bibitem{Cap80} D. M. Capper, D. R. T. Jones and P. van
Nieuwenhuizen, \np{B167}{479}{80} .
\bibitem{Sie80} W. Siegel, \pl{94B}{37}{80} .
\bibitem{Avd83} L. V. Avdeev and A. A. Vladimirov, \np{B219}{262}{83} .
\bibitem{Xu87} Z. Xu, D. Zhang and L. Chang, \np{B291}{392}{87} ;
R. Kleiss and W. J. Stirling, \np{B262}{235}{85} .
\bibitem{Gun85} J. Gunion and Z. Kunszt, \pl{B161}{333}{85} .
\bibitem{Pas79} G. Passarino and M. Veltman, \np{B160}{151}{79} .
\bibitem{Man91} M. L. Mangano and S. J. Parke, \prep{200}{301}{91} .
\bibitem{Dun92} Z. Bern and Dunbar, \np{B379}{562}{92} .
\bibitem{Abb81} L. F. Abbott, \np{B185}{189}{81} .
\bibitem{Cel79} W. Celmaster and R. J. Gonsalves, \pr{D20}{1420}{79} .
\bibitem{Alt81} G. Altarelli, G. Curci, G. Martinelli and S. Petrarca,
\np{B187}{461}{81} .
\bibitem{Wes74} J. Wess and B. Zumino, \np{B70}{39}{74} .
\bibitem{Par85} S. Parke and T. Taylor, \pl{B157}{81}{85} .
\bibitem{Kun86} Z. Kunszt, \np{B271}{333}{86} .
\bibitem{Gri77} M. T. Grisaru and H. N. Pendleton, \np{B124}{81}{77} .
\bibitem{Fur80} W. Furmanski and R. Petronzio, \pl{97B}{437}{80} .
\bibitem{Ant81} I. Antoniadis and E. G. Floratos, \np{B191}{217}{81} .
\bibitem{BDK92} Z. Bern, L. Dixon and D. A. Kosower, preprint
CERN-TH.6756/92.
\bibitem{'tH79} G. 't Hooft and M. Veltman, \np{B153}{365}{79} .
\bibitem{Ell81} R. K. Ellis, D. A. Ross and A. E. Terrano,
\np{B178}{421}{81} .
\end{thebibliography}
\end{document}